\hoffset -1 true cm
\overfullrule=0 pt
\raggedbottom
\newcount\equationno      \equationno=0
\newtoks\chapterno \xdef\chapterno{}
\newdimen\tabledimen  \tabledimen=\hsize
%
%
%

\def\eqname#1{\global \advance \equationno by 1 \relax
\xdef#1{{\noexpand{\rm}(\chapterno\number\equationno)}}#1}
%
%
%
%
\def\tablet#1#2{
\vbox{\tabskip=1em plus 4em minus 0.9em
\halign to #1{#2}} }
%
\def\tabmidrule{\noalign{\smallskip\hrule\smallskip}}             
\input epsf
%
%
%
%

\catcode `\@=11 

\def\@version{1.6}
\def\@verdate{18th September 1995}

%
%


\newif\ifprod@font

\ifx\@typeface\undefined
  \def\@typeface{Comp. Modern}\prod@fontfalse
\else
  \prod@fonttrue 
\fi

\def\newfam{\alloc@8\fam\chardef\sixt@@n} 

\ifprod@font
\font\fiverm=mtr10 at 5pt
\font\fivebf=mtbx10 at 5pt
\font\fiveit=mtti10 at 5pt
\font\fivesl=mtsl10 at 5pt
\font\fivett=cmtt8 at 5pt     \hyphenchar\fivett=-1
\font\fivecsc=mtcsc10 at 5pt
\font\fivesf=mtss10 at 5pt
\font\fivei=mtmi10 at 5pt      \skewchar\fivei='177
\font\fivesy=mtsy10 at 5pt     \skewchar\fivesy='60

\font\sixrm=mtr10 at 6pt
\font\sixbf=mtbx10 at 6pt
\font\sixit=mtti10 at 6pt
\font\sixsl=mtsl10 at 6pt
\font\sixtt=cmtt8 at 6pt      \hyphenchar\sixtt=-1
\font\sixcsc=mtcsc10 at 6pt
\font\sixsf=mtss10 at 6pt
\font\sixi=mtmi10 at 6pt       \skewchar\sixi='177
\font\sixsy=mtsy10 at 6pt      \skewchar\sixsy='60

\font\sevenrm=mtr10 at 7pt
\font\sevenbf=mtbx10 at 7pt
\font\sevenit=mtti10 at 7pt
\font\sevensl=mtsl10 at 7pt
\font\seventt=cmtt8 at 7pt     \hyphenchar\seventt=-1
\font\sevencsc=mtcsc10 at 7pt
\font\sevensf=mtss10 at 7pt
\font\seveni=mtmi10 at 7pt      \skewchar\seveni='177
\font\sevensy=mtsy10 at 7pt     \skewchar\sevensy='60

\font\eightrm=mtr10 at 8pt
\font\eightbf=mtbx10 at 8pt
\font\eightit=mtti10 at 8pt
\font\eighti=mtmi10 at 8pt      \skewchar\eighti='177
\font\eightsy=mtsy10 at 8pt     \skewchar\eightsy='60
\font\eightsl=mtsl10 at 8pt
\font\eighttt=cmtt8             \hyphenchar\eighttt=-1
\font\eightcsc=mtcsc10 at 8pt
\font\eightsf=mtss10 at 8pt

\font\ninerm=mtr10 at 9pt
\font\ninebf=mtbx10 at 9pt
\font\nineit=mtti10 at 9pt
\font\ninei=mtmi10 at 9pt      \skewchar\ninei='177
\font\ninesy=mtsy10 at 9pt     \skewchar\ninesy='60
\font\ninesl=mtsl10 at 9pt
\font\ninett=cmtt9             \hyphenchar\ninett=-1
\font\ninecsc=mtcsc10 at 9pt
\font\ninesf=mtss10 at 9pt

\font\tenrm=mtr10
\font\tenbf=mtbx10
\font\tenit=mtti10
\font\teni=mtmi10		\skewchar\teni='177
\font\tensy=mtsy10		\skewchar\tensy='60
\font\tenex=cmex10
\font\tensl=mtsl10
\font\tentt=cmtt10		\hyphenchar\tentt=-1
\font\tencsc=mtcsc10
\font\tensf=mtss10

\font\elevenrm=mtr10 at 11pt
\font\elevenbf=mtbx10 at 11pt
\font\elevenit=mtti10 at 11pt
\font\eleveni=mtmi10 at 11pt      \skewchar\eleveni='177
\font\elevensy=mtsy10 at 11pt     \skewchar\elevensy='60
\font\elevensl=mtsl10 at 11pt
\font\eleventt=cmtt10 at 11pt     \hyphenchar\eleventt=-1
\font\elevencsc=mtcsc10 at 11pt
\font\elevensf=mtss10 at 11pt

\font\twelverm=mtr10 at 12pt
\font\twelvebf=mtbx10 at 12pt
\font\twelveit=mtti10 at 12pt
\font\twelvesl=mtsl10 at 12pt
\font\twelvett=cmtt12             \hyphenchar\twelvett=-1
\font\twelvecsc=mtcsc10 at 12pt
\font\twelvesf=mtss10 at 12pt
\font\twelvei=mtmi10 at 12pt      \skewchar\twelvei='177
\font\twelvesy=mtsy10 at 12pt     \skewchar\twelvesy='60

\font\fourteenrm=mtr10 at 14pt
\font\fourteenbf=mtbx10 at 14pt
\font\fourteenit=mtti10 at 14pt
\font\fourteeni=mtmi10 at 14pt      \skewchar\fourteeni='177
\font\fourteensy=mtsy10 at 14pt     \skewchar\fourteensy='60
\font\fourteensl=mtsl10 at 14pt
\font\fourteentt=cmtt12 at 14pt     \hyphenchar\fourteentt=-1
\font\fourteencsc=mtcsc10 at 14pt
\font\fourteensf=mtss10 at 14pt

\font\seventeenrm=mtr10 at 17pt
\font\seventeenbf=mtbx10 at 17pt
\font\seventeenit=mtti10 at 17pt
\font\seventeeni=mtmi10 at 17pt      \skewchar\seventeeni='177
\font\seventeensy=mtsy10 at 17pt     \skewchar\seventeensy='60
\font\seventeensl=mtsl10 at 17pt
\font\seventeentt=cmtt12 at 17pt     \hyphenchar\seventeentt=-1
\font\seventeencsc=mtcsc10 at 17pt
\font\seventeensf=mtss10 at 17pt
\else
\font\fiverm=cmr5
\font\fivei=cmmi5             \skewchar\fivei='177
\font\fivesy=cmsy5            \skewchar\fivesy='60
\font\fivebf=cmbx5

\font\sixrm=cmr6
\font\sixi=cmmi6             \skewchar\sixi='177
\font\sixsy=cmsy6            \skewchar\sixsy='60
\font\sixbf=cmbx6

\font\sevenrm=cmr7
\font\sevenit=cmti7
\font\seveni=cmmi7             \skewchar\seveni='177
\font\sevensy=cmsy7            \skewchar\sevensy='60
\font\sevenbf=cmbx7

\font\eightrm=cmr8
\font\eightbf=cmbx8
\font\eightit=cmti8
\font\eighti=cmmi8			\skewchar\eighti='177
\font\eightsy=cmsy8			\skewchar\eightsy='60
\font\eightsl=cmsl8
\font\eighttt=cmtt8			\hyphenchar\eighttt=-1
\font\eightcsc=cmcsc10 at 8pt
\font\eightsf=cmss8

\font\ninerm=cmr9
\font\ninebf=cmbx9
\font\nineit=cmti9
\font\ninei=cmmi9			\skewchar\ninei='177
\font\ninesy=cmsy9			\skewchar\ninesy='60
\font\ninesl=cmsl9
\font\ninett=cmtt9			\hyphenchar\ninett=-1
\font\ninecsc=cmcsc10 at 9pt
\font\ninesf=cmss9

\font\tenrm=cmr10
\font\tenbf=cmbx10
\font\tenit=cmti10
\font\teni=cmmi10		\skewchar\teni='177
\font\tensy=cmsy10		\skewchar\tensy='60
\font\tenex=cmex10
\font\tensl=cmsl10
\font\tentt=cmtt10		\hyphenchar\tentt=-1
\font\tencsc=cmcsc10
\font\tensf=cmss10

\font\elevenrm=cmr10 scaled \magstephalf
\font\elevenbf=cmbx10 scaled \magstephalf
\font\elevenit=cmti10 scaled \magstephalf
\font\eleveni=cmmi10 scaled \magstephalf	\skewchar\eleveni='177
\font\elevensy=cmsy10 scaled \magstephalf	\skewchar\elevensy='60
\font\elevensl=cmsl10 scaled \magstephalf
\font\eleventt=cmtt10 scaled \magstephalf	\hyphenchar\eleventt=-1
\font\elevencsc=cmcsc10 scaled \magstephalf
\font\elevensf=cmss10 scaled \magstephalf

\font\twelverm=cmr10 scaled \magstep1
\font\twelvebf=cmbx10 scaled \magstep1
\font\twelvei=cmmi10 scaled \magstep1      \skewchar\twelvei='177
\font\twelvesy=cmsy10 scaled \magstep1     \skewchar\twelvesy='60

\font\fourteenrm=cmr10 scaled \magstep2
\font\fourteenbf=cmbx10 scaled \magstep2
\font\fourteenit=cmti10 scaled \magstep2
\font\fourteeni=cmmi10 scaled \magstep2		\skewchar\fourteeni='177
\font\fourteensy=cmsy10 scaled \magstep2	\skewchar\fourteensy='60
\font\fourteensl=cmsl10 scaled \magstep2
\font\fourteentt=cmtt10 scaled \magstep2	\hyphenchar\fourteentt=-1
\font\fourteencsc=cmcsc10 scaled \magstep2
\font\fourteensf=cmss10 scaled \magstep2

\font\seventeenrm=cmr10 scaled \magstep3
\font\seventeenbf=cmbx10 scaled \magstep3
\font\seventeenit=cmti10 scaled \magstep3
\font\seventeeni=cmmi10 scaled \magstep3	\skewchar\seventeeni='177
\font\seventeensy=cmsy10 scaled \magstep3	\skewchar\seventeensy='60
\font\seventeensl=cmsl10 scaled \magstep3
\font\seventeentt=cmtt10 scaled \magstep3	\hyphenchar\seventeentt=-1
\font\seventeencsc=cmcsc10 scaled \magstep3
\font\seventeensf=cmss10 scaled \magstep3
\fi

\def\hexnumber#1{\ifcase#1 0\or1\or2\or3\or4\or5\or6\or7\or8\or9\or
  A\or B\or C\or D\or E\or F\fi}

\def\makestrut{%
  \setbox\strutbox=\hbox{%
    \vrule height.7\baselineskip depth.3\baselineskip width \z@}%
}

\def\baselinestretch{1}
\newskip\tmp@bls

\def\b@ls#1{
  \tmp@bls=#1\relax
  \baselineskip=#1\relax\makestrut
  \normalbaselineskip=\baselinestretch\tmp@bls
  \normalbaselines
}

\def\nostb@ls#1{
  \normalbaselineskip=#1\relax
  \normalbaselines
  \makestrut
}

%

\newfam\scfam  
\newfam\sffam  

\def\mit{\fam\@ne}
\def\cal{\fam\tw@}
\def\em{\ifdim\fontdimen1\font>\z@ \rm\else\it\fi}

\textfont3=\tenex
\scriptfont3=\tenex
\scriptscriptfont3=\tenex

\setbox0=\hbox{\tenex B} \p@renwd=\wd0 

\def\eightpoint{
  \def\rm{\fam0\eightrm}%
  \textfont0=\eightrm \scriptfont0=\sixrm \scriptscriptfont0=\fiverm%
  \textfont1=\eighti  \scriptfont1=\sixi  \scriptscriptfont1=\fivei%
  \textfont2=\eightsy \scriptfont2=\sixsy \scriptscriptfont2=\fivesy%
  \textfont\itfam=\eightit\def\it{\fam\itfam\eightit}%
  \ifprod@font
    \scriptfont\itfam=\sixit
      \scriptscriptfont\itfam=\fiveit
  \else
    \scriptfont\itfam=\eightit
      \scriptscriptfont\itfam=\eightit
  \fi
  \textfont\bffam=\eightbf%
    \scriptfont\bffam=\sixbf%
      \scriptscriptfont\bffam=\fivebf%
  \def\bf{\fam\bffam\eightbf}%
  \textfont\slfam=\eightsl\def\sl{\fam\slfam\eightsl}%
  \ifprod@font
    \scriptfont\slfam=\sixsl
      \scriptscriptfont\slfam=\fivesl
  \else
    \scriptfont\slfam=\eightsl
      \scriptscriptfont\slfam=\eightsl
  \fi
  \textfont\ttfam=\eighttt\def\tt{\fam\ttfam\eighttt}%
  \ifprod@font
    \scriptfont\ttfam=\sixtt
      \scriptscriptfont\ttfam=\fivett
  \else
    \scriptfont\ttfam=\eighttt
      \scriptscriptfont\ttfam=\eighttt
  \fi
  \textfont\scfam=\eightcsc\def\sc{\fam\scfam\eightcsc}%
  \ifprod@font
    \scriptfont\scfam=\sixcsc
      \scriptscriptfont\scfam=\fivecsc
  \else
    \scriptfont\scfam=\eightcsc
      \scriptscriptfont\scfam=\eightcsc
  \fi
  \textfont\sffam=\eightsf\def\sf{\fam\sffam\eightsf}%
  \ifprod@font
    \scriptfont\sffam=\sixsf
      \scriptscriptfont\sffam=\fivesf
  \else
    \scriptfont\sffam=\eightsf
      \scriptscriptfont\sffam=\eightsf
  \fi
  \def\oldstyle{\fam\@ne\eighti}%
  \b@ls{10pt}\rm\@viiipt%
}
\def\@viiipt{}

\def\ninepoint{
  \def\rm{\fam0\ninerm}%
  \textfont0=\ninerm \scriptfont0=\sixrm \scriptscriptfont0=\fiverm%
  \textfont1=\ninei  \scriptfont1=\sixi  \scriptscriptfont1=\fivei%
  \textfont2=\ninesy \scriptfont2=\sixsy \scriptscriptfont2=\fivesy%
  \textfont\itfam=\nineit\def\it{\fam\itfam\nineit}%
  \ifprod@font
    \scriptfont\itfam=\sixit
      \scriptscriptfont\itfam=\fiveit
  \else
    \scriptfont\itfam=\nineit
      \scriptscriptfont\itfam=\nineit
  \fi
  \textfont\bffam=\ninebf%
    \scriptfont\bffam=\sixbf%
      \scriptscriptfont\bffam=\fivebf%
  \def\bf{\fam\bffam\ninebf}%
  \textfont\slfam=\ninesl\def\sl{\fam\slfam\ninesl}%
  \ifprod@font
    \scriptfont\slfam=\sixsl
      \scriptscriptfont\slfam=\fivesl
  \else
    \scriptfont\slfam=\ninesl
      \scriptscriptfont\slfam=\ninesl
  \fi
  \textfont\ttfam=\ninett\def\tt{\fam\ttfam\ninett}%
  \ifprod@font
    \scriptfont\ttfam=\sixtt
      \scriptscriptfont\ttfam=\fivett
  \else
    \scriptfont\ttfam=\ninett
      \scriptscriptfont\ttfam=\ninett
  \fi
  \textfont\scfam=\ninecsc\def\sc{\fam\scfam\ninecsc}%
  \ifprod@font
    \scriptfont\scfam=\sixcsc
      \scriptscriptfont\scfam=\fivecsc
  \else
    \scriptfont\scfam=\ninecsc
      \scriptscriptfont\scfam=\ninecsc
  \fi
  \textfont\sffam=\ninesf\def\sf{\fam\sffam\ninesf}%
  \ifprod@font
    \scriptfont\sffam=\sixsf
      \scriptscriptfont\sffam=\fivesf
  \else
    \scriptfont\sffam=\ninesf
      \scriptscriptfont\sffam=\ninesf
  \fi
  \def\oldstyle{\fam\@ne\ninei}%
  \b@ls{\TextLeading plus \Feathering}\rm\@ixpt%
}
\def\@ixpt{}

\def\tenpoint{
  \def\rm{\fam0\tenrm}%
  \textfont0=\tenrm \scriptfont0=\sevenrm \scriptscriptfont0=\fiverm%
  \textfont1=\teni  \scriptfont1=\seveni  \scriptscriptfont1=\fivei%
  \textfont2=\tensy \scriptfont2=\sevensy \scriptscriptfont2=\fivesy%
  \textfont\itfam=\tenit\def\it{\fam\itfam\tenit}%
  \ifprod@font
    \scriptfont\itfam=\sevenit
      \scriptscriptfont\itfam=\fiveit
  \else
    \scriptfont\itfam=\tenit
      \scriptscriptfont\itfam=\tenit
  \fi
  \textfont\bffam=\tenbf%
    \scriptfont\bffam=\sevenbf%
      \scriptscriptfont\bffam=\fivebf%
  \def\bf{\fam\bffam\tenbf}%
  \textfont\slfam=\tensl\def\sl{\fam\slfam\tensl}%
  \ifprod@font
    \scriptfont\slfam=\sevensl
      \scriptscriptfont\slfam=\fivesl
  \else
    \scriptfont\slfam=\tensl
      \scriptscriptfont\slfam=\tensl
  \fi
  \textfont\ttfam=\tentt\def\tt{\fam\ttfam\tentt}%
  \ifprod@font
    \scriptfont\ttfam=\seventt
      \scriptscriptfont\ttfam=\fivett
  \else
    \scriptfont\ttfam=\tentt
      \scriptscriptfont\ttfam=\tentt
  \fi
  \textfont\scfam=\tencsc\def\sc{\fam\scfam\tencsc}%
  \ifprod@font
    \scriptfont\scfam=\sevencsc
      \scriptscriptfont\scfam=\fivecsc
  \else
    \scriptfont\scfam=\tencsc
      \scriptscriptfont\scfam=\tencsc
  \fi
  \textfont\sffam=\tensf\def\sf{\fam\sffam\tensf}%
  \ifprod@font
    \scriptfont\sffam=\sevensf
      \scriptscriptfont\sffam=\fivesf
  \else
    \scriptfont\sffam=\tensf
      \scriptscriptfont\sffam=\tensf
  \fi
  \def\oldstyle{\fam\@ne\teni}%
  \b@ls{11pt}\rm\@xpt%
}
\def\@xpt{}

\def\elevenpoint{
  \def\rm{\fam0\elevenrm}%
  \textfont0=\elevenrm \scriptfont0=\eightrm \scriptscriptfont0=\sixrm%
  \textfont1=\eleveni  \scriptfont1=\eighti  \scriptscriptfont1=\sixi%
  \textfont2=\elevensy \scriptfont2=\eightsy \scriptscriptfont2=\sixsy%
  \textfont\itfam=\elevenit\def\it{\fam\itfam\elevenit}%
  \ifprod@font
    \scriptfont\itfam=\eightit
      \scriptscriptfont\itfam=\sixit
  \else
    \scriptfont\itfam=\elevenit
      \scriptscriptfont\itfam=\elevenit
  \fi
  \textfont\bffam=\elevenbf%
    \scriptfont\bffam=\eightbf%
      \scriptscriptfont\bffam=\sixbf%
  \def\bf{\fam\bffam\elevenbf}%
  \textfont\slfam=\elevensl\def\sl{\fam\slfam\elevensl}%
  \ifprod@font
    \scriptfont\slfam=\eightsl
      \scriptscriptfont\slfam=\sixsl
  \else
    \scriptfont\slfam=\elevensl
      \scriptscriptfont\slfam=\elevensl
  \fi
  \textfont\ttfam=\eleventt\def\tt{\fam\ttfam\eleventt}%
  \ifprod@font
    \scriptfont\ttfam=\eighttt
      \scriptscriptfont\ttfam=\sixtt
  \else
    \scriptfont\ttfam=\eleventt
      \scriptscriptfont\ttfam=\eleventt
  \fi
  \textfont\scfam=\elevencsc\def\sc{\fam\scfam\elevencsc}%
  \ifprod@font
    \scriptfont\scfam=\eightcsc
      \scriptscriptfont\scfam=\sixcsc
  \else
    \scriptfont\scfam=\elevencsc
      \scriptscriptfont\scfam=\elevencsc
  \fi
  \textfont\sffam=\elevensf\def\sf{\fam\sffam\elevensf}%
  \ifprod@font
    \scriptfont\sffam=\eightsf
      \scriptscriptfont\sffam=\sixsf
  \else
    \scriptfont\sffam=\elevensf
      \scriptscriptfont\sffam=\elevensf
  \fi
  \def\oldstyle{\fam\@ne\eleveni}%
  \b@ls{13pt}\rm\@xipt%
}
\def\@xipt{}

\def\fourteenpoint{
  \def\rm{\fam0\fourteenrm}%
  \textfont0\fourteenrm  \scriptfont0\tenrm  \scriptscriptfont0\sevenrm%
  \textfont1\fourteeni   \scriptfont1\teni   \scriptscriptfont1\seveni%
  \textfont2\fourteensy  \scriptfont2\tensy  \scriptscriptfont2\sevensy%
  \textfont\itfam=\fourteenit\def\it{\fam\itfam\fourteenit}%
  \ifprod@font
    \scriptfont\itfam=\tenit
      \scriptscriptfont\itfam=\sevenit
  \else
    \scriptfont\itfam=\fourteenit
      \scriptscriptfont\itfam=\fourteenit
  \fi
  \textfont\bffam=\fourteenbf%
    \scriptfont\bffam=\tenbf%
      \scriptscriptfont\bffam=\sevenbf%
  \def\bf{\fam\bffam\fourteenbf}%
  \textfont\slfam=\fourteensl\def\sl{\fam\slfam\fourteensl}%
  \ifprod@font
    \scriptfont\slfam=\tensl
      \scriptscriptfont\slfam=\sevensl
  \else
    \scriptfont\slfam=\fourteensl
      \scriptscriptfont\slfam=\fourteensl
  \fi
  \textfont\ttfam=\fourteentt\def\tt{\fam\ttfam\fourteentt}%
  \ifprod@font
    \scriptfont\ttfam=\tentt
      \scriptscriptfont\ttfam=\seventt
  \else
    \scriptfont\ttfam=\fourteentt
      \scriptscriptfont\ttfam=\fourteentt
  \fi
  \textfont\scfam=\fourteencsc\def\sc{\fam\scfam\fourteencsc}%
  \ifprod@font
    \scriptfont\scfam=\tencsc
      \scriptscriptfont\scfam=\sevencsc
  \else
    \scriptfont\scfam=\fourteencsc
      \scriptscriptfont\scfam=\fourteencsc
  \fi
  \textfont\sffam=\fourteensf\def\sf{\fam\sffam\fourteensf}%
  \ifprod@font
    \scriptfont\sffam=\tensf
      \scriptscriptfont\sffam=\sevensf
  \else
    \scriptfont\sffam=\fourteensf
      \scriptscriptfont\sffam=\fourteensf
  \fi
  \def\oldstyle{\fam\@ne\fourteeni}%
  \b@ls{17pt}\rm\@xivpt%
}
\def\@xivpt{}

\def\seventeenpoint{
  \def\rm{\fam0\seventeenrm}%
  \textfont0\seventeenrm  \scriptfont0\twelverm  \scriptscriptfont0\tenrm%
  \textfont1\seventeeni   \scriptfont1\twelvei   \scriptscriptfont1\teni%
  \textfont2\seventeensy  \scriptfont2\twelvesy  \scriptscriptfont2\tensy%
  \textfont\itfam=\seventeenit\def\it{\fam\itfam\seventeenit}%
  \ifprod@font
    \scriptfont\itfam=\twelveit
      \scriptscriptfont\itfam=\tenit
  \else
    \scriptfont\itfam=\seventeenit
      \scriptscriptfont\itfam=\seventeenit
  \fi
  \textfont\bffam=\seventeenbf%
    \scriptfont\bffam=\twelvebf%
      \scriptscriptfont\bffam=\tenbf%
  \def\bf{\fam\bffam\seventeenbf}%
  \textfont\slfam=\seventeensl\def\sl{\fam\slfam\seventeensl}%
  \ifprod@font
    \scriptfont\slfam=\twelvesl
      \scriptscriptfont\slfam=\tensl
  \else
    \scriptfont\slfam=\seventeensl
      \scriptscriptfont\slfam=\seventeensl
  \fi
  \textfont\ttfam=\seventeentt\def\tt{\fam\ttfam\seventeentt}%
  \ifprod@font
    \scriptfont\ttfam=\twelvett
      \scriptscriptfont\ttfam=\tentt
  \else
    \scriptfont\ttfam=\seventeentt
      \scriptscriptfont\ttfam=\seventeentt
  \fi
  \textfont\scfam=\seventeencsc\def\sc{\fam\scfam\seventeencsc}%
  \ifprod@font
    \scriptfont\scfam=\twelvecsc
      \scriptscriptfont\scfam=\tencsc
  \else
    \scriptfont\scfam=\seventeencsc
      \scriptscriptfont\scfam=\seventeencsc
  \fi
  \textfont\sffam=\seventeensf\def\sf{\fam\sffam\seventeensf}%
  \ifprod@font
    \scriptfont\sffam=\twelvesf
      \scriptscriptfont\sffam=\tensf
  \else
    \scriptfont\sffam=\seventeensf
      \scriptscriptfont\sffam=\seventeensf
  \fi
  \def\oldstyle{\fam\@ne\seventeeni}%
  \b@ls{20pt}\rm\@xviipt%
}
\def\@xviipt{}

\lineskip=1pt      \normallineskip=\lineskip
\lineskiplimit=\z@ \normallineskiplimit=\lineskiplimit


\def\,{\relax\ifmmode \mskip\thinmuskip\else \thinspace\fi}
\let\protect=\relax

\long\def\@ifundefined#1#2#3{\expandafter\ifx\csname
  #1\endcsname\relax#2\else#3\fi}




\newtoks\math@groups \math@groups={}
\def\addtom@thgroup#1#2{#1\expandafter{\the#1#2}} 



\def\addtosizeh@ok#1#2#3#4{%
  \expandafter\def\csname @#1pt\endcsname{%
    \def\s@ze{#2}\def\ss@ze{#3}\def\sss@ze{#4}\the\math@groups%
  }%
}



\let\resetsizehook=\addtosizeh@ok


\ifprod@font
  \addtosizeh@ok{viii} {8} {6}  {5}
  \addtosizeh@ok{ix}   {9} {6}  {5}
  \addtosizeh@ok{x}    {10}{7}  {5}
  \addtosizeh@ok{xi}   {11}{8}  {6}
  \addtosizeh@ok{xiv}  {14}{10} {7}
  \addtosizeh@ok{xvii} {17}{12}{10}
\else
  \addtosizeh@ok{viii} {8}     {6}     {5}
  \addtosizeh@ok{ix}   {9}     {6}     {5}
  \addtosizeh@ok{x}    {10}    {7}     {5}
  \addtosizeh@ok{xi}   {10.95} {8}     {6}
  \addtosizeh@ok{xiv}  {14.4}  {10}    {7}
  \addtosizeh@ok{xvii} {17.28} {12}    {10}
\fi

\def\get@font#1#2#3{%
  \edef\fonts@ze{\romannumeral#3}
  \edef\fontn@me{\fonts@ze#1}
  \@ifundefined{\fontn@me}%
    {
     \global\expandafter\font\csname \fontn@me\endcsname=#2 at #3pt}%
    {}%
}

\def\ass@tfont#1#2{%
  \xdef\fam@name{\csname #1\endcsname}%
  \xdef\font@name{\csname #2\endcsname}%
  \let\textfont@name\font@name
  \textfont\fam@name\textfont@name
}

\def\ass@sfont#1#2{%
  \xdef\fam@name{\csname #1\endcsname}%
  \xdef\font@name{\csname #2\endcsname}%
  \let\textfont@name\font@name
  \scriptfont\fam@name\textfont@name
}

\def\ass@ssfont#1#2{%
  \xdef\fam@name{\csname #1\endcsname}%
  \xdef\font@name{\csname #2\endcsname}%
  \let\textfont@name\font@name
  \scriptscriptfont\fam@name\textfont@name
}


\def\NewSymbolFont#1#2{%
  \expandafter\ifx\csname sym#1fam\endcsname\relax 
    \expandafter\newfam\csname sym#1fam\endcsname
    \expandafter\edef\csname sym#1fam\endcsname{\the\allocationnumber}%
    \addtom@thgroup\math@groups{%
      \get@font{#1}{#2}{\s@ze}%
      \ass@tfont{sym#1fam}{\fontn@me}%
      \get@font{#1}{#2}{\ss@ze}%
      \ass@sfont{sym#1fam}{\fontn@me}%
      \get@font{#1}{#2}{\sss@ze}%
      \ass@ssfont{sym#1fam}{\fontn@me}%
    }%
  \else
    \errmessage{Family `#1' already defined}%
  \fi
}


\def\NewMathSymbol#1#2#3#4{%
  \edef\f@mly{\expandafter\hexnumber{\csname sym#3fam\endcsname}}%
  \mathchardef#1="#2\f@mly#4\relax
}


\newif\ifd@f

\def\NewMathDelimiter#1#2#3#4#5#6{%
  \d@ftrue
  \expandafter\ifx\csname sym#3fam\endcsname\relax
    \d@ffalse \errmessage{Family `#3' is not defined}%
  \fi
  \expandafter\ifx\csname sym#5fam\endcsname\relax
    \d@ffalse \errmessage{Family `#5' is not defined}%
  \fi
  \ifd@f
    \edef\f@mly{\expandafter\hexnumber{\csname sym#3fam\endcsname}}%
    \edef\f@mlytw@{\expandafter\hexnumber{\csname sym#5fam\endcsname}}%
    \xdef#1{\delimiter"#2\f@mly #4\f@mlytw@ #6\relax}%
  \fi
}


\def\setboxz@h{\setbox\z@\hbox}
\def\wdz@{\wd\z@}
\def\boxz@{\box\z@}
\def\setbox@ne{\setbox\@ne}
\def\wd@ne{\wd\@ne}

\def\math@atom#1#2{%
   \binrel@{#1}\binrel@@{#2}}
\def\binrel@#1{\setboxz@h{\thinmuskip0mu
  \medmuskip\m@ne mu\thickmuskip\@ne mu$#1\m@th$}%
 \setbox@ne\hbox{\thinmuskip0mu\medmuskip\m@ne mu\thickmuskip
  \@ne mu${}#1{}\m@th$}%
 \setbox\tw@\hbox{\hskip\wd@ne\hskip-\wdz@}}
\def\binrel@@#1{\ifdim\wd2<\z@\mathbin{#1}\else\ifdim\wd\tw@>\z@
 \mathrel{#1}\else{#1}\fi\fi}

\def\m@thit{1}

\def\set@skchar#1{\global\expandafter\skewchar
  \csname\fontn@me\endcsname=#1\relax}

\def\NewMathAlphabet#1#2#3{%
  \def\tst{#3}%
  \ifx\tst\empty\else 
    \expandafter\gdef\csname #1@sc\endcsname{}
  \fi
  \expandafter\def\csname #1\endcsname{
    \protect\csname @#1\endcsname}%
  \expandafter\def\csname @#1\endcsname##1{
    {%
    \begingroup
      \get@font{#1}{#2}{\s@ze}%
      \@ifundefined{#1@sc}{}{\set@skchar{#3}}%
      \ass@tfont{m@thit}{\fontn@me}%
      \get@font{#1}{#2}{\ss@ze}%
      \@ifundefined{#1@sc}{}{\set@skchar{#3}}%
      \ass@sfont{m@thit}{\fontn@me}%
      \get@font{#1}{#2}{\sss@ze}%
      \@ifundefined{#1@sc}{}{\set@skchar{#3}}%
      \ass@ssfont{m@thit}{\fontn@me}%
      \math@atom{##1}{%
      \mathchoice%
        {\hbox{$\m@th\displaystyle##1$}}%
        {\hbox{$\m@th\textstyle##1$}}%
        {\hbox{$\m@th\scriptstyle##1$}}%
        {\hbox{$\m@th\scriptscriptstyle##1$}}}%
    \endgroup
    }%
  }%
}


\newif\iffirstta  \firsttatrue

\def\set@hchar#1{\global\expandafter\hyphenchar
  \csname\fontn@me\endcsname=#1\relax}

\def\NewTextAlphabet#1#2#3{%
  \iffirstta
    \global\firsttafalse
    \newfam\scratchfam
    \edef\scrt@fam{\the\allocationnumber}%
  \fi
  \def\tst{#3}%
  \ifx\tst\empty\else 
    \expandafter\gdef\csname #1@hc\endcsname{}
  \fi
  \expandafter\def\csname #1\endcsname{
    \protect\csname t@#1\endcsname}%
  \long\expandafter\def\csname t@#1\endcsname##1{
    \ifmmode
      \typeout{Warning: do not use \expandafter\string\csname #1\endcsname
        \space in math mode}\fi%
    {%
      \get@font{#1}{#2}{\s@ze}\let\t@xtfnt=\fontn@me\relax
      \@ifundefined{#1@hc}{}{\set@hchar{#3}}%
      \ass@tfont{scrt@fam}{\fontn@me}%
      \get@font{#1}{#2}{\ss@ze}%
      \@ifundefined{#1@hc}{}{\set@hchar{#3}}%
      \ass@sfont{scrt@fam}{\fontn@me}%
      \get@font{#1}{#2}{\sss@ze}%
      \@ifundefined{#1@hc}{}{\set@hchar{#3}}%
      \ass@ssfont{scrt@fam}{\fontn@me}%
      \fam\scratchfam\csname\t@xtfnt\endcsname
    ##1%
    }%
  }%
  \expandafter\def\csname #1shape
    \endcsname{\protect\csname @#1shape\endcsname}%
  \expandafter\def\csname @#1shape\endcsname{
    \ifmmode
      \typeout{Warning: do not use \expandafter\string\csname
        #1shape\endcsname \space in math mode}\fi
      \get@font{#1}{#2}{\s@ze}\let\t@xtfnt=\fontn@me\relax
      \@ifundefined{#1@hc}{}{\set@hchar{#3}}%
      \ass@tfont{scrt@fam}{\fontn@me}%
      \get@font{#1}{#2}{\ss@ze}%
      \@ifundefined{#1@hc}{}{\set@hchar{#3}}%
      \ass@sfont{scrt@fam}{\fontn@me}%
      \get@font{#1}{#2}{\sss@ze}%
      \@ifundefined{#1@hc}{}{\set@hchar{#3}}%
      \ass@ssfont{scrt@fam}{\fontn@me}%
      \fam\scratchfam\csname\t@xtfnt\endcsname
  }%
}


\ifprod@font
  \def\math@itfnt{mtmib10}
  \def\math@syfnt{mtbsy10}
\else
  \def\math@itfnt{cmmib10}
  \def\math@syfnt{cmbsy10}
\fi

\def\m@thsy{2}

\def\bmath{\protect\@bmath}
\def\@bmath#1{%
  {%
  \begingroup
    \get@font{mthit}{\math@itfnt}{\s@ze}\set@skchar{'177}%
    \ass@tfont{m@thit}{\fontn@me}%
    \get@font{mthit}{\math@itfnt}{\ss@ze}\set@skchar{'177}%
    \ass@sfont{m@thit}{\fontn@me}%
    \get@font{mthit}{\math@itfnt}{\sss@ze}\set@skchar{'177}%
    \ass@ssfont{m@thit}{\fontn@me}%
    \get@font{mthsy}{\math@syfnt}{\s@ze}\set@skchar{'60}%
    \ass@tfont{m@thsy}{\fontn@me}%
    \get@font{mthsy}{\math@syfnt}{\ss@ze}\set@skchar{'60}%
    \ass@sfont{m@thsy}{\fontn@me}%
    \get@font{mthsy}{\math@syfnt}{\sss@ze}\set@skchar{'60}%
    \ass@ssfont{m@thsy}{\fontn@me}%
    \math@atom{#1}{%
    \mathchoice%
      {\hbox{$\m@th\displaystyle#1$}}%
      {\hbox{$\m@th\textstyle#1$}}%
      {\hbox{$\m@th\scriptstyle#1$}}%
      {\hbox{$\m@th\scriptscriptstyle#1$}}}%
  \endgroup
  }%
}



\def\diameter{{\ifmmode\mathchoice
{\ooalign{\hfil\hbox{$\displaystyle/$}\hfil\crcr
{\hbox{$\displaystyle\mathchar"20D$}}}}
{\ooalign{\hfil\hbox{$\textstyle/$}\hfil\crcr
{\hbox{$\textstyle\mathchar"20D$}}}}
{\ooalign{\hfil\hbox{$\scriptstyle/$}\hfil\crcr
{\hbox{$\scriptstyle\mathchar"20D$}}}}
{\ooalign{\hfil\hbox{$\scriptscriptstyle/$}\hfil\crcr
{\hbox{$\scriptscriptstyle\mathchar"20D$}}}}
\else{\ooalign{\hfil/\hfil\crcr\mathhexbox20D}}%
\fi}}

\def\sq{\ifmmode\squareforqed\else{\unskip\nobreak\hfil
\penalty50\hskip1em\null\nobreak\hfil\squareforqed
\parfillskip=0pt\finalhyphendemerits=0\endgraf}\fi}
\def\squareforqed{\hbox{\rlap{$\sqcap$}$\sqcup$}}


\def\bbbc{{\mathchoice {\setbox0=\hbox{$\displaystyle\rm C$}\hbox{\hbox
to0pt{\kern0.4\wd0\vrule height0.9\ht0\hss}\box0}}
{\setbox0=\hbox{$\textstyle\rm C$}\hbox{\hbox
to0pt{\kern0.4\wd0\vrule height0.9\ht0\hss}\box0}}
{\setbox0=\hbox{$\scriptstyle\rm C$}\hbox{\hbox
to0pt{\kern0.4\wd0\vrule height0.9\ht0\hss}\box0}}
{\setbox0=\hbox{$\scriptscriptstyle\rm C$}\hbox{\hbox
to0pt{\kern0.4\wd0\vrule height0.9\ht0\hss}\box0}}}}
\def\bbbq{{\mathchoice {\setbox0=\hbox{$\displaystyle\rm
Q$}\hbox{\raise
0.15\ht0\hbox to0pt{\kern0.4\wd0\vrule height0.8\ht0\hss}\box0}}
{\setbox0=\hbox{$\textstyle\rm Q$}\hbox{\raise
0.15\ht0\hbox to0pt{\kern0.4\wd0\vrule height0.8\ht0\hss}\box0}}
{\setbox0=\hbox{$\scriptstyle\rm Q$}\hbox{\raise
0.15\ht0\hbox to0pt{\kern0.4\wd0\vrule height0.7\ht0\hss}\box0}}
{\setbox0=\hbox{$\scriptscriptstyle\rm Q$}\hbox{\raise
0.15\ht0\hbox to0pt{\kern0.4\wd0\vrule height0.7\ht0\hss}\box0}}}}
\def\bbbt{{\mathchoice {\setbox0=\hbox{$\displaystyle\rm
T$}\hbox{\hbox to0pt{\kern0.3\wd0\vrule height0.9\ht0\hss}\box0}}
{\setbox0=\hbox{$\textstyle\rm T$}\hbox{\hbox
to0pt{\kern0.3\wd0\vrule height0.9\ht0\hss}\box0}}
{\setbox0=\hbox{$\scriptstyle\rm T$}\hbox{\hbox
to0pt{\kern0.3\wd0\vrule height0.9\ht0\hss}\box0}}
{\setbox0=\hbox{$\scriptscriptstyle\rm T$}\hbox{\hbox
to0pt{\kern0.3\wd0\vrule height0.9\ht0\hss}\box0}}}}
\def\bbbs{{\mathchoice
{\setbox0=\hbox{$\displaystyle     \rm S$}\hbox{\raise0.5\ht0\hbox
to0pt{\kern0.35\wd0\vrule height0.45\ht0\hss}\hbox
to0pt{\kern0.55\wd0\vrule height0.5\ht0\hss}\box0}}
{\setbox0=\hbox{$\textstyle        \rm S$}\hbox{\raise0.5\ht0\hbox
to0pt{\kern0.35\wd0\vrule height0.45\ht0\hss}\hbox
to0pt{\kern0.55\wd0\vrule height0.5\ht0\hss}\box0}}
{\setbox0=\hbox{$\scriptstyle      \rm S$}\hbox{\raise0.5\ht0\hbox
to0pt{\kern0.35\wd0\vrule height0.45\ht0\hss}\raise0.05\ht0\hbox
to0pt{\kern0.5\wd0\vrule height0.45\ht0\hss}\box0}}
{\setbox0=\hbox{$\scriptscriptstyle\rm S$}\hbox{\raise0.5\ht0\hbox
to0pt{\kern0.4\wd0\vrule height0.45\ht0\hss}\raise0.05\ht0\hbox
to0pt{\kern0.55\wd0\vrule height0.45\ht0\hss}\box0}}}}
\def\bbbz{{\mathchoice {\hbox{$\sf\textstyle Z\kern-0.4em Z$}}
{\hbox{$\sf\textstyle Z\kern-0.4em Z$}}
{\hbox{$\sf\scriptstyle Z\kern-0.3em Z$}}
{\hbox{$\sf\scriptscriptstyle Z\kern-0.2em Z$}}}}


\def\Nulle{0} 
\def\Afe{1}   
\def\Hae{2}   
\def\Hbe{3}   
\def\Hce{4}   
\def\Hde{5}   


\newcount\LastMac       \LastMac=\Nulle

\newskip\half      \half=5.5pt plus 1.5pt minus 2.25pt
\newskip\one       \one=11pt plus 3pt minus 5.5pt
\newskip\onehalf   \onehalf=16.5pt plus 5.5pt minus 8.25pt
\newskip\two       \two=22pt plus 5.5pt minus 11pt

\def\Half{\addvspace{\half}}
\def\One{\addvspace{\one}}
\def\OneHalf{\addvspace{\onehalf}}
\def\Two{\addvspace{\two}}

\def\Raggedright{
  \rightskip=\z@ plus \hsize\relax
}

\def\Fullout{
  \rightskip=\z@\relax
}

\def\Hang#1#2{
  \hangindent=#1%
  \hangafter=#2\relax
}


\newif\ifsp@page
\def\pagestyle#1{\csname ps@#1\endcsname}
\def\thispagestyle#1{\global\sp@pagetrue\gdef\sp@type{#1}}

\def\ps@titlepage{%
  \def\@oddhead{\eightpoint\noindent \the\CatchLine
    \ifprod@font\else\qquad Printed\ \today\qquad
      (MN plain \TeX\ macros\ v\@version)\fi \hfil}%
  \let\@evenhead=\@oddhead
  \def\@oddfoot{\eightpoint\copyright\ \@pubyear\ RAS\hfil}%
  \def\@evenfoot{\hfil\eightpoint\noindent\copyright\ \@pubyear\ RAS}%
}

\def\ps@headings{%
  \def\@oddhead{\elevenpoint\it\noindent
    \hfill\the\RightHeader\hskip1.5em\rm\folio}%
  \def\@evenhead{\elevenpoint\noindent
    \folio\hskip1.5em\it\the\LeftHeader\hfill}%
  \def\@oddfoot{\eightpoint\noindent\copyright\ \@pubyear\ RAS,
    MNRAS {\bf \@volume}, \@pagerange\hfil}%
  \def\@evenfoot{\hfil\eightpoint\copyright\ \@pubyear\ RAS,
    MNRAS {\bf \@volume}, \@pagerange}%
}

\def\ps@plate{%
  \def\@oddhead{\eightpoint\noindent\plt@cap\hfil}%
  \def\@evenhead{\eightpoint\noindent\plt@cap\hfil}%
  \def\@oddfoot{\eightpoint\noindent\copyright\ \@pubyear\ RAS,
    MNRAS {\bf \@volume}, \@pagerange\hfil}%
  \def\@evenfoot{\hfil\eightpoint\copyright\ \@pubyear\ RAS,
    MNRAS {\bf \@volume}, \@pagerange}%
}



\def\title#1{
  \bgroup
    \vbox to 8pt{\vss}%
    \seventeenpoint
    \Raggedright
    \noindent \strut{\bf #1}\par
  \egroup
}

\def\author#1{
  \bgroup
    \ifnum\LastMac=\Afe \OneHalf\else \vskip 21pt\fi
    \fourteenpoint
    \Raggedright
    \noindent \strut #1\par
    \vskip 3pt%
  \egroup
}

\def\affiliation#1{
  \bgroup
    \vskip -4pt%
    \eightpoint
    \Raggedright
    \noindent \strut {\it #1}\par
  \egroup
  \LastMac=\Afe\relax
}

\def\acceptedline#1{
  \bgroup
    \Two
    \eightpoint
    \Raggedright
    \noindent \strut #1\par
  \egroup
}

\long\def\abstract#1{%
  \bgroup
    \vskip 20pt%
    \leftskip 11pc\rightskip\z@
    \noindent{\ninebf ABSTRACT}\par
    \tenpoint
    \Fullout
    \noindent #1\par
  \egroup
}

\long\def\keywords#1{
  \bgroup
    \Half
    \leftskip 11pc\rightskip\z@
    \tenpoint
    \Fullout
    \noindent\hbox{\bf Key words:}\ #1\par
  \egroup
}


\def\maketitle{%
  \EndOpening
  \ifsinglecol \else \MakePage\fi
}



\def\@nameuse#1{\csname #1\endcsname}
\def\arabic#1{\@arabic{\@nameuse{#1}}}
\def\alph#1{\@alph{\@nameuse{#1}}}
\def\Alph#1{\@Alph{\@nameuse{#1}}}
\def\@arabic#1{\number #1}
\def\@Alph#1{\ifcase#1\or A\or B\or C\or D\else\@Ialph{#1}\fi}
\def\@Ialph#1{\ifcase#1\or \or \or \or \or E\or F\or G\or H\or I\or J\or
   K\or L\or M\or N\or O\or P\or Q\or R\or S\or T\or U\or V\or W\or X\or
   Y\or Z\else\errmessage{Counter out of range}\fi}
\def\@alph#1{\ifcase#1\or a\or b\or c\or d\else\@ialph{#1}\fi}
\def\@ialph#1{\ifcase#1\or \or \or \or \or e\or f\or g\or h\or i\or j\or
   k\or l\or m\or n\or o\or p\or q\or r\or s\or t\or u\or v\or w\or x\or y\or
   z\else\errmessage{Counter out of range}\fi}


\newcount\Eqnno
\newcount\SubEqnno

\def\theeq{\arabic{Eqnno}}
\def\thesubeq{\alph{SubEqnno}}

\def\stepeq{\relax
  \global\SubEqnno \z@
  \global\advance\Eqnno \@ne\relax
  {\rm (\theeq)}%
}

\def\startsubeq{\relax
  \global\SubEqnno \z@
  \global\advance\Eqnno \@ne\relax
  \stepsubeq
}

\def\stepsubeq{\relax
  \global\advance\SubEqnno \@ne\relax
  {\rm (\theeq\thesubeq)}%
}


\newcount\Sec        
\newcount\SecSec
\newcount\SecSecSec

\def\thesection{\arabic{Sec}}
\def\thesubsection{\thesection.\arabic{SecSec}}
\def\thesubsubsection{\thesubsection.\arabic{SecSecSec}}

\Sec=\z@

\def\:{\let\@sptoken= } \:  
\def\:{\@xifnch} \expandafter\def\: {\futurelet\@tempc\@ifnch}

\def\@ifnextchar#1#2#3{%
  \let\@tempMACe #1%
  \def\@tempMACa{#2}%
  \def\@tempMACb{#3}%
  \futurelet \@tempMACc\@ifnch%
}

\def\@ifnch{%
\ifx \@tempMACc \@sptoken%
  \let\@tempMACd\@xifnch%
\else%
  \ifx \@tempMACc \@tempMACe%
    \let\@tempMACd\@tempMACa%
  \else%
    \let\@tempMACd\@tempMACb%
  \fi%
\fi%
\@tempMACd%
}

\def\@ifstar#1#2{\@ifnextchar *{\def\@tempMACa*{#1}\@tempMACa}{#2}}

\newskip\@tempskipb

\def\addvspace#1{%
  \ifvmode\else \endgraf\fi%
  \ifdim\lastskip=\z@%
    \vskip #1\relax%
  \else%
    \@tempskipb#1\relax\@xaddvskip%
  \fi%
}

\def\@xaddvskip{%
  \ifdim\lastskip<\@tempskipb%
    \vskip-\lastskip%
    \vskip\@tempskipb\relax%
  \else%
    \ifdim\@tempskipb<\z@%
      \ifdim\lastskip<\z@ \else%
        \advance\@tempskipb\lastskip%
        \vskip-\lastskip\vskip\@tempskipb%
      \fi%
    \fi%
  \fi%
}

\newskip\@tmpSKIP

\def\addpen#1{%
  \ifvmode
    \if@nobreak
    \else
      \ifdim\lastskip=\z@
        \penalty#1\relax
      \else
        \@tmpSKIP=\lastskip
        \vskip -\lastskip
        \penalty#1\vskip\@tmpSKIP
      \fi
    \fi
  \fi
}

\newcount\@clubpen   \@clubpen=\clubpenalty
\newif\if@nobreak    \@nobreakfalse

\def\@noafterindent{%
  \global\@nobreaktrue
  \everypar{\if@nobreak
              \global\@nobreakfalse
              \clubpenalty \@M
              {\setbox\z@\lastbox}%
              \LastMac=\Nulle\relax%
            \else
              \clubpenalty \@clubpen
              \everypar{}%
            \fi}%
}

\newcount\gds@cbrk   \gds@cbrk=-300

\def\@nohdbrk{\interlinepenalty \@M\relax}

\let\@par=\par
\def\@restorepar{\def\par{\@par}}

\newif\if@endpe   \@endpefalse
 
\def\@doendpe{\@endpetrue \@nobreakfalse \LastMac=\Nulle\relax%
     \def\par{\@restorepar\everypar{}\par\@endpefalse}%
              \everypar{\setbox\z@\lastbox\everypar{}\@endpefalse}%
}

\def\section{\@ifstar{\@ssection}{\@section}}

\def\@section#1{
  \if@nobreak
    \everypar{}%
    \ifnum\LastMac=\Hae \addvspace{\half}\fi
  \else
    \addpen{\gds@cbrk}%
    \addvspace{\two}%
  \fi
  \bgroup
    \ninepoint\bf
    \Raggedright
    \global\advance\Sec \@ne
    \ifappendix
      \global\Eqnno=\z@ \global\SubEqnno=\z@\relax
      \def\ch@ck{#1}%
      \ifx\ch@ck\empty \def\c@lon{}\else\def\c@lon{:}\fi
      \noindent\@nohdbrk APPENDIX\ \thesection\c@lon\hskip 0.5em%
        \uppercase{#1}\par
    \else
      \noindent\@nohdbrk\thesection\hskip 1pc \uppercase{#1}\par
    \fi
    \global\SecSec=\z@
  \egroup
  \nobreak
  \vskip\half
  \nobreak
  \@noafterindent
  \LastMac=\Hae\relax
}

\def\@ssection#1{
  \if@nobreak
    \everypar{}%
    \ifnum\LastMac=\Hae \addvspace{\half}\fi
  \else
    \addpen{\gds@cbrk}%
    \addvspace{\two}%
  \fi
  \bgroup
    \ninepoint\bf
    \Raggedright
    \noindent\@nohdbrk\uppercase{#1}\par
  \egroup
  \nobreak
  \vskip\half
  \nobreak
  \@noafterindent
  \LastMac=\Hae\relax
}

\def\subsection{\@ifstar{\@ssubsection}{\@subsection}}

\def\@subsection#1{
  \if@nobreak
    \everypar{}%
    \ifnum\LastMac=\Hae \addvspace{1pt plus 1pt minus .5pt}\fi
  \else
    \addpen{\gds@cbrk}%
    \addvspace{\onehalf}%
  \fi
  \bgroup
    \ninepoint\bf
    \Raggedright
    \global\advance\SecSec \@ne
    \noindent\@nohdbrk\thesubsection \hskip 1pc\relax #1\par
    \global\SecSecSec=\z@
  \egroup
  \nobreak
  \vskip\half
  \nobreak
  \@noafterindent
  \LastMac=\Hbe\relax
}

\def\@ssubsection#1{
  \if@nobreak
    \everypar{}%
    \ifnum\LastMac=\Hae \addvspace{1pt plus 1pt minus .5pt}\fi
  \else
    \addpen{\gds@cbrk}%
    \addvspace{\onehalf}%
  \fi
  \bgroup
    \ninepoint\bf
    \Raggedright
    \noindent\@nohdbrk #1\par
  \egroup
  \nobreak
  \vskip\half
  \nobreak
  \@noafterindent
  \LastMac=\Hbe\relax
}

\def\subsubsection{\@ifstar{\@ssubsubsection}{\@subsubsection}}

\def\@subsubsection#1{
  \if@nobreak
    \everypar{}%
    \ifnum\LastMac=\Hbe \addvspace{1pt plus 1pt minus .5pt}\fi
  \else
    \addpen{\gds@cbrk}%
    \addvspace{\onehalf}%
  \fi
  \bgroup
    \ninepoint\it
    \Raggedright
    \global\advance\SecSecSec \@ne
    \noindent\@nohdbrk\thesubsubsection \hskip 1pc\relax #1\par
  \egroup
  \nobreak
  \vskip\half
  \nobreak
  \@noafterindent
  \LastMac=\Hce\relax
}

\def\@ssubsubsection#1{
  \if@nobreak
    \everypar{}%
    \ifnum\LastMac=\Hbe \addvspace{1pt plus 1pt minus .5pt}\fi
  \else
    \addpen{\gds@cbrk}%
    \addvspace{\onehalf}%
  \fi
  \bgroup
    \ninepoint\it
    \Raggedright
    \noindent\@nohdbrk #1\par
  \egroup
  \nobreak
  \vskip\half
  \nobreak
  \@noafterindent
  \LastMac=\Hce\relax
}

\def\paragraph#1{
  \if@nobreak
    \everypar{}%
  \else
    \addpen{\gds@cbrk}%
    \addvspace{\one}%
  \fi%
  \bgroup%
    \ninepoint\it
    \noindent #1\ \nobreak%
  \egroup
  \LastMac=\Hde\relax
  \ignorespaces
}


\newif\ifappendix

\def\appendix{%
  \global\appendixtrue
  \def\thesection{\Alph{Sec}}%
  \def\thesubsection{\thesection\arabic{SecSec}}%
  \def\theeq{\thesection\arabic{Eqnno}}%
  \Sec=\z@ \SecSec=\z@ \SecSecSec=\z@ \Eqnno=\z@ \SubEqnno=\z@\relax
}




\def\beginlist{%
  \par\if@nobreak \else\addvspace{\half}\fi%
  \bgroup%
    \ninepoint
    \let\item=\list@item%
}

\def\list@item{%
  \par\noindent\hskip 1em\relax%
  \ignorespaces%
}

\def\endlist{\par\egroup\addvspace{\half}\@doendpe}


\def\beginrefs{%
  \par
  \bgroup
    \eightpoint
    \Fullout
    \let\bibitem=\bib@item
}

\def\bib@item{%
  \par\parindent=1.5em\Hang{1.5em}{1}%
  \everypar={\Hang{1.5em}{1}\ignorespaces}%
  \noindent\ignorespaces
}

\def\endrefs{\par\egroup\@doendpe}


\newtoks\CatchLine

\def\@journal{Mon.\ Not.\ R.\ Astron.\ Soc.\ }  
\def\@pubyear{1994}        
\def\@pagerange{000--000}  
\def\@volume{000}          
\def\@microfiche{}         %

\def\pubyear#1{\gdef\@pubyear{#1}\@makecatchline}
\def\pagerange#1{\gdef\@pagerange{#1}\@makecatchline}
\def\volume#1{\gdef\@volume{#1}\@makecatchline}
\def\microfiche#1{\gdef\@microfiche{and Microfiche\ #1}\@makecatchline}

\def\@makecatchline{%
  \global\CatchLine{%
    {\rm \@journal {\bf \@volume},\ \@pagerange\ (\@pubyear)\ \@microfiche}}%
}

\@makecatchline 

\newtoks\LeftHeader

\newtoks\RightHeader

\def\PageHead{
  \begingroup
    \ifsp@page
      \csname ps@\sp@type\endcsname
    \fi
    \ifodd\pageno
      \let\the@head=\@oddhead
    \else
      \let\the@head=\@evenhead
    \fi
    \vbox to \z@{\vskip-22.5\p@%
      \hbox to \PageWidth{\vbox to8.5\p@{}%
        \the@head
      }%
    \vss}%
  \endgroup
  \nointerlineskip
}

\gdef\PageFoot{%
  \nointerlineskip%
  \begingroup
  \ifsp@page
    \csname ps@\sp@type\endcsname
    \global\sp@pagefalse
  \fi
  \vbox to 22pt{\vfil%
    \hbox to \PageWidth{%
      \eightpoint\strut\noindent
      \ifodd\pageno
        \@oddfoot
      \else
        \@evenfoot
      \fi
    }%
  }%
  \endgroup
}

\def\today{%
  \number\day\space
  \ifcase\month\or January\or February\or March\or April\or May\or June\or
    July\or August\or September\or October\or November\or December\fi
  \space\number\year%
}

\def\authorcomment#1{%
  \gdef\PageFoot{%
    \nointerlineskip%
    \vbox to 20pt{\vfil%
      \hbox to \PageWidth{\elevenpoint\noindent \hfil #1 \hfil}}%
  }%
}


\newif\ifplate@page
\newbox\plt@box

\def\beginplatepage{%
  \let\plate=\plate@head
  \let\caption=\fig@caption
  \global\setbox\plt@box=\vbox\bgroup
  \TEMPDIMEN=\PageWidth 
  \hsize=\PageWidth\relax
}

\def\endplatepage{\par\egroup\global\plate@pagetrue}
\def\plate@head#1{\gdef\plt@cap{#1}}


\def\letters{%
  \gdef\folio{\ifnum\pageno<\z@ L\romannumeral-\pageno
    \else L\number\pageno \fi}%
}


\newdimen\mathindent

\global\mathindent=\z@
\global\everydisplay{\global\@dspwd=\displaywidth\displaysetup}


\def\@displaylines#1{
  {}$\displ@y\hbox{\vbox{\halign{$\@lign\hfil\displaystyle##\hfil$\crcr
  #1\crcr}}}${}%
}

\def\@eqalign#1{\null\vcenter{\openup\jot\m@th
  \ialign{\strut\hfil$\displaystyle{##}$&$\displaystyle{{}##}$\hfil
      \crcr#1\crcr}}%
}

\def\@eqalignno#1{
  \global\advance\@dspwd by -\mathindent%
  {}$\displ@y\hbox{\vbox{\halign to\@dspwd%
  {\hfil$\@lign\displaystyle{##}$\tabskip\z@skip
  &$\@lign\displaystyle{{}##}$\hfil\tabskip\centering
  &\llap{$\@lign##$}\tabskip\z@skip\crcr
  #1\crcr}}}${}%
}


\global\let\displaylines=\@displaylines
\global\let\eqalign=\@eqalign
\global\let\eqalignno=\@eqalignno
\global\let\leqalignno=\@eqalignno

\newdimen\@dspwd   \@dspwd=\z@
\newif\if@eqno
\newif\if@leqno
\newtoks\@eqn
\newtoks\@eq

\def\displaysetup#1$${\displaytest#1\eqno\eqno\displaytest}

\def\displaytest#1\eqno#2\eqno#3\displaytest{%
 \if!#3!\ldisplaytest#1\leqno\leqno\ldisplaytest
 \else\@eqnotrue\@leqnofalse\@eqn={#2}\@eq={#1}\fi
 \generaldisplay$$}

\def\ldisplaytest#1\leqno#2\leqno#3\ldisplaytest{%
\@eq={#1}%
 \if!#3!\@eqnofalse\else\@eqnotrue\@leqnotrue
  \@eqn={#2}\fi}

\def\generaldisplay{%
  \if@eqno
    \if@leqno
      \hbox to \displaywidth{\noindent
        \rlap{$\displaystyle\the\@eqn$}%
        \hskip\mathindent$\displaystyle\the\@eq$\hfil}%
    \else
      \hbox to \displaywidth{\noindent
        \hskip\mathindent
        $\displaystyle\the\@eq$\hfil$\displaystyle\the\@eqn$}%
    \fi
  \else
    \hbox to \displaywidth{\noindent
      \hskip\mathindent$\displaystyle\the\@eq$\hfil}%
  \fi
}


\def\@notice{%
  \par\Two%
  \noindent{\b@ls{11pt}\ninerm This paper has been produced using the
    Royal Astronomical Society/Blackwell Science \TeX\ macros.\par}%
}

\outer\def\bye{\@notice\par\vfill\supereject\end}


\def\start@mess{%
  Monthly notices of the RAS journal style (\@typeface)\space
    v\@version,\space \@verdate.%
}

\everyjob{\Warn{\start@mess}}



\newif\if@debug \@debugfalse  

\def\Print#1{\if@debug\immediate\write16{#1}\else \fi}
\def\Warn#1{\immediate\write16{#1}}
\def\wlog#1{}

\newcount\Iteration 

\def\Single{0} \def\Double{1}                 
\def\Figure{0} \def\Table{1}                  

\def\InStack{0}  
\def\InZoneA{1}
\def\InZoneB{2}
\def\InZoneC{3}

\newcount\TEMPCOUNT 
\newdimen\TEMPDIMEN 
\newbox\TEMPBOX     
\newbox\VOIDBOX     

\newcount\LengthOfStack 
\newcount\MaxItems      
\newcount\StackPointer
\newcount\Point         
\newcount\NextFigure    
\newcount\NextTable     
\newcount\NextItem      

\newcount\StatusStack   
\newcount\NumStack      
\newcount\TypeStack     
\newcount\SpanStack     
\newcount\BoxStack      

\newcount\ItemSTATUS    
\newcount\ItemNUMBER    
\newcount\ItemTYPE      
\newcount\ItemSPAN      
\newbox\ItemBOX         
\newdimen\ItemSIZE      

\newdimen\PageHeight    
\newdimen\TextLeading   
\newdimen\Feathering    
\newcount\LinesPerPage  
\newdimen\ColumnWidth   
\newdimen\ColumnGap     
\newdimen\PageWidth     
\newdimen\BodgeHeight   
\newcount\Leading       

\newdimen\ZoneBSize  
\newdimen\TextSize   
\newbox\ZoneABOX     
\newbox\ZoneBBOX     
\newbox\ZoneCBOX     

\newif\ifFirstSingleItem
\newif\ifFirstZoneA
\newif\ifMakePageInComplete
\newif\ifMoreFigures \MoreFiguresfalse 
\newif\ifMoreTables  \MoreTablesfalse  

\newif\ifFigInZoneB 
\newif\ifFigInZoneC 
\newif\ifTabInZoneB 
\newif\ifTabInZoneC

\newif\ifZoneAFullPage

\newbox\MidBOX    
\newbox\LeftBOX
\newbox\RightBOX
\newbox\PageBOX   

\newif\ifLeftCOL  
\LeftCOLtrue

\newdimen\ZoneBAdjust

\newcount\ItemFits
\def\Yes{1}
\def\No{2}


\MaxItems=15
\NextFigure=\z@        
\NextTable=\@ne

\BodgeHeight=6pt
\TextLeading=11pt    
\Leading=11
\Feathering=\z@      
\LinesPerPage=61     
\topskip=\TextLeading
\ColumnWidth=20pc    
\ColumnGap=2pc       

\newskip\ItemSepamount  
\ItemSepamount=\TextLeading plus \TextLeading minus 4pt

\parskip=\z@ plus .1pt
\parindent=18pt
\widowpenalty=\z@
\clubpenalty=10000
\tolerance=1500
\hbadness=1500
\abovedisplayskip=6pt plus 2pt minus 1pt
\belowdisplayskip=6pt plus 2pt minus 1pt
\abovedisplayshortskip=6pt plus 2pt minus 1pt
\belowdisplayshortskip=6pt plus 2pt minus 1pt

\frenchspacing

\ninepoint 

\PageHeight=682pt
\PageWidth=2\ColumnWidth
\advance\PageWidth by \ColumnGap

\pagestyle{headings}




\newcount\DUMMY \StatusStack=\allocationnumber
\newcount\DUMMY \newcount\DUMMY \newcount\DUMMY 
\newcount\DUMMY \newcount\DUMMY \newcount\DUMMY 
\newcount\DUMMY \newcount\DUMMY \newcount\DUMMY
\newcount\DUMMY \newcount\DUMMY \newcount\DUMMY 
\newcount\DUMMY \newcount\DUMMY \newcount\DUMMY

\newcount\DUMMY \NumStack=\allocationnumber
\newcount\DUMMY \newcount\DUMMY \newcount\DUMMY 
\newcount\DUMMY \newcount\DUMMY \newcount\DUMMY 
\newcount\DUMMY \newcount\DUMMY \newcount\DUMMY 
\newcount\DUMMY \newcount\DUMMY \newcount\DUMMY 
\newcount\DUMMY \newcount\DUMMY \newcount\DUMMY

\newcount\DUMMY \TypeStack=\allocationnumber
\newcount\DUMMY \newcount\DUMMY \newcount\DUMMY 
\newcount\DUMMY \newcount\DUMMY \newcount\DUMMY 
\newcount\DUMMY \newcount\DUMMY \newcount\DUMMY 
\newcount\DUMMY \newcount\DUMMY \newcount\DUMMY 
\newcount\DUMMY \newcount\DUMMY \newcount\DUMMY

\newcount\DUMMY \SpanStack=\allocationnumber
\newcount\DUMMY \newcount\DUMMY \newcount\DUMMY 
\newcount\DUMMY \newcount\DUMMY \newcount\DUMMY 
\newcount\DUMMY \newcount\DUMMY \newcount\DUMMY 
\newcount\DUMMY \newcount\DUMMY \newcount\DUMMY 
\newcount\DUMMY \newcount\DUMMY \newcount\DUMMY

\newbox\DUMMY   \BoxStack=\allocationnumber
\newbox\DUMMY   \newbox\DUMMY \newbox\DUMMY 
\newbox\DUMMY   \newbox\DUMMY \newbox\DUMMY 
\newbox\DUMMY   \newbox\DUMMY \newbox\DUMMY 
\newbox\DUMMY   \newbox\DUMMY \newbox\DUMMY 
\newbox\DUMMY   \newbox\DUMMY \newbox\DUMMY

\def\wlog{\immediate\write\m@ne}


\def\GetItemAll#1{%
 \GetItemSTATUS{#1}
 \GetItemNUMBER{#1}
 \GetItemTYPE{#1}
 \GetItemSPAN{#1}
 \GetItemBOX{#1}
}

\def\GetItemSTATUS#1{%
 \Point=\StatusStack
 \advance\Point by #1
 \global\ItemSTATUS=\count\Point
}

\def\GetItemNUMBER#1{%
 \Point=\NumStack
 \advance\Point by #1
 \global\ItemNUMBER=\count\Point
}

\def\GetItemTYPE#1{%
 \Point=\TypeStack
 \advance\Point by #1
 \global\ItemTYPE=\count\Point
}

\def\GetItemSPAN#1{%
 \Point\SpanStack
 \advance\Point by #1
 \global\ItemSPAN=\count\Point
}

\def\GetItemBOX#1{%
 \Point=\BoxStack
 \advance\Point by #1
 \global\setbox\ItemBOX=\vbox{\copy\Point}
 \global\ItemSIZE=\ht\ItemBOX
 \global\advance\ItemSIZE by \dp\ItemBOX
 \TEMPCOUNT=\ItemSIZE
 \divide\TEMPCOUNT by \Leading
 \divide\TEMPCOUNT by 65536
 \advance\TEMPCOUNT \@ne
 \ItemSIZE=\TEMPCOUNT pt
 \global\multiply\ItemSIZE by \Leading
}


\def\JoinStack{%
 \ifnum\LengthOfStack=\MaxItems 
  \Warn{WARNING: Stack is full...some items will be lost!}
 \else
  \Point=\StatusStack
  \advance\Point by \LengthOfStack
  \global\count\Point=\ItemSTATUS
  \Point=\NumStack
  \advance\Point by \LengthOfStack
  \global\count\Point=\ItemNUMBER
  \Point=\TypeStack
  \advance\Point by \LengthOfStack
  \global\count\Point=\ItemTYPE
  \Point\SpanStack
  \advance\Point by \LengthOfStack
  \global\count\Point=\ItemSPAN
  \Point=\BoxStack
  \advance\Point by \LengthOfStack
  \global\setbox\Point=\vbox{\copy\ItemBOX}
  \global\advance\LengthOfStack \@ne
  \ifnum\ItemTYPE=\Figure 
   \global\MoreFigurestrue
  \else
   \global\MoreTablestrue
  \fi
 \fi
}


\def\LeaveStack#1{%
 {\Iteration=#1
 \loop
 \ifnum\Iteration<\LengthOfStack
  \advance\Iteration \@ne
  \GetItemSTATUS{\Iteration}
   \advance\Point by \m@ne
   \global\count\Point=\ItemSTATUS
  \GetItemNUMBER{\Iteration}
   \advance\Point by \m@ne
   \global\count\Point=\ItemNUMBER
  \GetItemTYPE{\Iteration}
   \advance\Point by \m@ne
   \global\count\Point=\ItemTYPE
  \GetItemSPAN{\Iteration}
   \advance\Point by \m@ne
   \global\count\Point=\ItemSPAN
  \GetItemBOX{\Iteration}
   \advance\Point by \m@ne
   \global\setbox\Point=\vbox{\copy\ItemBOX}
 \repeat}
 \global\advance\LengthOfStack by \m@ne
}


\newif\ifStackNotClean

\def\CleanStack{%
 \StackNotCleantrue
 {\Iteration=\z@
  \loop
   \ifStackNotClean
    \GetItemSTATUS{\Iteration}
    \ifnum\ItemSTATUS=\InStack
     \advance\Iteration \@ne
     \else
      \LeaveStack{\Iteration}
    \fi
   \ifnum\LengthOfStack<\Iteration
    \StackNotCleanfalse
   \fi
 \repeat}
}


\def\FindItem#1#2{%
 \global\StackPointer=\m@ne 
 {\Iteration=\z@
  \loop
  \ifnum\Iteration<\LengthOfStack
   \GetItemSTATUS{\Iteration}
   \ifnum\ItemSTATUS=\InStack
    \GetItemTYPE{\Iteration}
    \ifnum\ItemTYPE=#1
     \GetItemNUMBER{\Iteration}
     \ifnum\ItemNUMBER=#2
      \global\StackPointer=\Iteration
      \Iteration=\LengthOfStack 
     \fi
    \fi
   \fi
  \advance\Iteration \@ne
 \repeat}
}


\def\FindNext{%
 \global\StackPointer=\m@ne 
 {\Iteration=\z@
  \loop
  \ifnum\Iteration<\LengthOfStack
   \GetItemSTATUS{\Iteration}
   \ifnum\ItemSTATUS=\InStack
    \GetItemTYPE{\Iteration}
   \ifnum\ItemTYPE=\Figure
    \ifMoreFigures
      \global\NextItem=\Figure
      \global\StackPointer=\Iteration
      \Iteration=\LengthOfStack 
    \fi
   \fi
   \ifnum\ItemTYPE=\Table
    \ifMoreTables
      \global\NextItem=\Table
      \global\StackPointer=\Iteration
      \Iteration=\LengthOfStack 
    \fi
   \fi
  \fi
  \advance\Iteration \@ne
 \repeat}
}


\def\ChangeStatus#1#2{%
 \Point=\StatusStack
 \advance\Point by #1
 \global\count\Point=#2
}



\def\Zone{\InZoneA}

\ZoneBAdjust=\z@

\def\MakePage{
 \global\ZoneBSize=\PageHeight
 \global\TextSize=\ZoneBSize
 \global\ZoneAFullPagefalse
 \global\topskip=\TextLeading
 \MakePageInCompletetrue
 \MoreFigurestrue
 \MoreTablestrue
 \FigInZoneBfalse
 \FigInZoneCfalse
 \TabInZoneBfalse
 \TabInZoneCfalse
 \global\FirstSingleItemtrue
 \global\FirstZoneAtrue
 \global\setbox\ZoneABOX=\box\VOIDBOX
 \global\setbox\ZoneBBOX=\box\VOIDBOX
 \global\setbox\ZoneCBOX=\box\VOIDBOX
 \loop
  \ifMakePageInComplete
 \FindNext
 \ifnum\StackPointer=\m@ne
  \NextItem=\m@ne
  \MoreFiguresfalse
  \MoreTablesfalse
 \fi
 \ifnum\NextItem=\Figure
   \FindItem{\Figure}{\NextFigure}
   \ifnum\StackPointer=\m@ne \global\MoreFiguresfalse
   \else
    \GetItemSPAN{\StackPointer}
    \ifnum\ItemSPAN=\Single \def\Zone{\InZoneB}\relax
     \ifFigInZoneC \global\MoreFiguresfalse\fi
    \else
     \def\Zone{\InZoneA}
     \ifFigInZoneB \def\Zone{\InZoneC}\fi
    \fi
   \fi
   \ifMoreFigures\Print{}\FigureItems\fi
 \fi
\ifnum\NextItem=\Table
   \FindItem{\Table}{\NextTable}
   \ifnum\StackPointer=\m@ne \global\MoreTablesfalse
   \else
    \GetItemSPAN{\StackPointer}
    \ifnum\ItemSPAN=\Single\relax
     \ifTabInZoneC \global\MoreTablesfalse\fi
    \else
     \def\Zone{\InZoneA}
     \ifTabInZoneB \def\Zone{\InZoneC}\fi
    \fi
   \fi
   \ifMoreTables\Print{}\TableItems\fi
 \fi
   \MakePageInCompletefalse 
   \ifMoreFigures\MakePageInCompletetrue\fi
   \ifMoreTables\MakePageInCompletetrue\fi
 \repeat
 \ifZoneAFullPage
  \global\TextSize=\z@
  \global\ZoneBSize=\z@
  \global\vsize=\z@\relax
  \global\topskip=\z@\relax
  \vbox to \z@{\vss}
  \eject
 \else
 \global\advance\ZoneBSize by -\ZoneBAdjust
 \global\vsize=\ZoneBSize
 \global\hsize=\ColumnWidth
 \global\ZoneBAdjust=\z@
 \ifdim\TextSize<23pt
 \Warn{}
 \Warn{* Making column fall short: TextSize=\the\TextSize *}
 \vskip-\lastskip\eject\fi
 \fi
}

\def\MakeRightCol{
 \global\TextSize=\ZoneBSize
 \MakePageInCompletetrue
 \MoreFigurestrue
 \MoreTablestrue
 \global\FirstSingleItemtrue
 \global\setbox\ZoneBBOX=\box\VOIDBOX
 \def\Zone{\InZoneB}
 \loop
  \ifMakePageInComplete
 \FindNext
 \ifnum\StackPointer=\m@ne
  \NextItem=\m@ne
  \MoreFiguresfalse
  \MoreTablesfalse
 \fi
 \ifnum\NextItem=\Figure
   \FindItem{\Figure}{\NextFigure}
   \ifnum\StackPointer=\m@ne \MoreFiguresfalse
   \else
    \GetItemSPAN{\StackPointer}
    \ifnum\ItemSPAN=\Double\relax
     \MoreFiguresfalse\fi
   \fi
   \ifMoreFigures\Print{}\FigureItems\fi
 \fi
 \ifnum\NextItem=\Table
   \FindItem{\Table}{\NextTable}
   \ifnum\StackPointer=\m@ne \MoreTablesfalse
   \else
    \GetItemSPAN{\StackPointer}
    \ifnum\ItemSPAN=\Double\relax
     \MoreTablesfalse\fi
   \fi
   \ifMoreTables\Print{}\TableItems\fi
 \fi
   \MakePageInCompletefalse 
   \ifMoreFigures\MakePageInCompletetrue\fi
   \ifMoreTables\MakePageInCompletetrue\fi
 \repeat
 \ifZoneAFullPage
  \global\TextSize=\z@
  \global\ZoneBSize=\z@
  \global\vsize=\z@\relax
  \global\topskip=\z@\relax
  \vbox to \z@{\vss}
  \eject
 \else
 \global\vsize=\ZoneBSize
 \global\hsize=\ColumnWidth
 \ifdim\TextSize<23pt
 \Warn{}
 \Warn{* Making column fall short: TextSize=\the\TextSize *}
 \vskip-\lastskip\eject\fi
\fi
}

\def\FigureItems{
 \Print{Considering...}
 \ShowItem{\StackPointer}
 \GetItemBOX{\StackPointer} 
 \GetItemSPAN{\StackPointer}
  \CheckFitInZone 
  \ifnum\ItemFits=\Yes
   \ifnum\ItemSPAN=\Single
     \ChangeStatus{\StackPointer}{\InZoneB} 
     \global\FigInZoneBtrue
     \ifFirstSingleItem
      \hbox{}\vskip-\BodgeHeight
     \global\advance\ItemSIZE by \TextLeading
     \fi
     \unvbox\ItemBOX\ItemSep
     \global\FirstSingleItemfalse
     \global\advance\TextSize by -\ItemSIZE
     \global\advance\TextSize by -\TextLeading
   \else
    \ifFirstZoneA
     \global\advance\ItemSIZE by \TextLeading
     \global\FirstZoneAfalse\fi
    \global\advance\TextSize by -\ItemSIZE
    \global\advance\TextSize by -\TextLeading
    \global\advance\ZoneBSize by -\ItemSIZE
    \global\advance\ZoneBSize by -\TextLeading
    \ifFigInZoneB\relax
     \else
     \ifdim\TextSize<3\TextLeading
     \global\ZoneAFullPagetrue
     \fi
    \fi
    \ChangeStatus{\StackPointer}{\Zone}
    \ifnum\Zone=\InZoneC \global\FigInZoneCtrue\fi
  \fi
   \Print{TextSize=\the\TextSize}
   \Print{ZoneBSize=\the\ZoneBSize}
  \global\advance\NextFigure \@ne
   \Print{This figure has been placed.}
  \else
   \Print{No space available for this figure...holding over.}
   \Print{}
   \global\MoreFiguresfalse
  \fi
}

\def\TableItems{
 \Print{Considering...}
 \ShowItem{\StackPointer}
 \GetItemBOX{\StackPointer} 
 \GetItemSPAN{\StackPointer}
  \CheckFitInZone 
  \ifnum\ItemFits=\Yes
   \ifnum\ItemSPAN=\Single
    \ChangeStatus{\StackPointer}{\InZoneB}
     \global\TabInZoneBtrue
     \ifFirstSingleItem
      \hbox{}\vskip-\BodgeHeight
     \global\advance\ItemSIZE by \TextLeading
     \fi
     \unvbox\ItemBOX\ItemSep
     \global\FirstSingleItemfalse
     \global\advance\TextSize by -\ItemSIZE
     \global\advance\TextSize by -\TextLeading
   \else
    \ifFirstZoneA
    \global\advance\ItemSIZE by \TextLeading
    \global\FirstZoneAfalse\fi
    \global\advance\TextSize by -\ItemSIZE
    \global\advance\TextSize by -\TextLeading
    \global\advance\ZoneBSize by -\ItemSIZE
    \global\advance\ZoneBSize by -\TextLeading
    \ifFigInZoneB\relax
     \else
     \ifdim\TextSize<3\TextLeading
     \global\ZoneAFullPagetrue
     \fi
    \fi
    \ChangeStatus{\StackPointer}{\Zone}
    \ifnum\Zone=\InZoneC \global\TabInZoneCtrue\fi
   \fi
  \global\advance\NextTable \@ne
   \Print{This table has been placed.}
  \else
  \Print{No space available for this table...holding over.}
   \Print{}
   \global\MoreTablesfalse
  \fi
}


\def\CheckFitInZone{%
{\advance\TextSize by -\ItemSIZE
 \advance\TextSize by -\TextLeading
 \ifFirstSingleItem
  \advance\TextSize by \TextLeading
 \fi
 \ifnum\Zone=\InZoneA\relax
  \else \advance\TextSize by -\ZoneBAdjust
 \fi
 \ifdim\TextSize<3\TextLeading \global\ItemFits=\No
 \else \global\ItemFits=\Yes\fi}
}

\def\BeginOpening{%
  \ninepoint
  \thispagestyle{titlepage}%
  \global\setbox\ItemBOX=\vbox\bgroup%
    \hsize=\PageWidth%
    \hrule height \z@
    \ifsinglecol\vskip 6pt\fi 
}

\let\begintopmatter=\BeginOpening  

\def\EndOpening{%
  \One
  \egroup
  \ifsinglecol
    \box\ItemBOX%
    \vskip\TextLeading plus 2\TextLeading
    \@noafterindent
  \else
    \ItemNUMBER=\z@%
    \ItemTYPE=\Figure
    \ItemSPAN=\Double
    \ItemSTATUS=\InStack
    \JoinStack
  \fi
}


\newif\if@here  \@herefalse

\def\no@float{\global\@heretrue}
\let\nofloat=\relax 

\def\beginfigure{%
  \@ifstar{\global\@dfloattrue \@bfigure}{\global\@dfloatfalse \@bfigure}%
}

\def\@bfigure#1{%
  \par
  \if@dfloat
    \ItemSPAN=\Double
    \TEMPDIMEN=\PageWidth
  \else
    \ItemSPAN=\Single
    \TEMPDIMEN=\ColumnWidth
  \fi
  \ifsinglecol
    \TEMPDIMEN=\PageWidth
  \else
    \ItemSTATUS=\InStack
    \ItemNUMBER=#1%
    \ItemTYPE=\Figure
  \fi
  \bgroup
    \hsize=\TEMPDIMEN
    \global\setbox\ItemBOX=\vbox\bgroup
      \eightpoint\nostb@ls{10pt}%
      \let\caption=\fig@caption
      \ifsinglecol \let\nofloat=\no@float\fi
}

\def\fig@caption#1{%
  \vskip 5.5pt plus 6pt%
  \bgroup 
    \eightpoint\nostb@ls{10pt}%
    \setbox\TEMPBOX=\hbox{#1}%
    \ifdim\wd\TEMPBOX>\TEMPDIMEN
      \noindent \unhbox\TEMPBOX\par
    \else
      \hbox to \hsize{\hfil\unhbox\TEMPBOX\hfil}%
    \fi
  \egroup
}

\def\endfigure{%
  \par\egroup 
  \egroup
  \ifsinglecol
    \if@here \midinsert\global\@herefalse\else \topinsert\fi
      \unvbox\ItemBOX
    \endinsert
  \else
    \JoinStack
    \Print{Processing source for figure \the\ItemNUMBER}%
  \fi
}


\newbox\tab@cap@box
\def\tab@caption#1{\global\setbox\tab@cap@box=\hbox{#1\par}}

\newtoks\tab@txt@toks
\long\def\tab@txt#1{\global\tab@txt@toks={#1}\global\table@txttrue}

\newif\iftable@txt  \table@txtfalse
\newif\if@dfloat    \@dfloatfalse

\def\begintable{%
  \@ifstar{\global\@dfloattrue \@btable}{\global\@dfloatfalse \@btable}%
}

\def\@btable#1{%
  \par
  \if@dfloat
    \ItemSPAN=\Double
    \TEMPDIMEN=\PageWidth
  \else
    \ItemSPAN=\Single
    \TEMPDIMEN=\ColumnWidth
  \fi
  \ifsinglecol
    \TEMPDIMEN=\PageWidth
  \else
    \ItemSTATUS=\InStack
    \ItemNUMBER=#1%
    \ItemTYPE=\Table
  \fi
  \bgroup
    \eightpoint\nostb@ls{10pt}%
    \global\setbox\ItemBOX=\vbox\bgroup
      \let\caption=\tab@caption
      \let\tabletext=\tab@txt
      \ifsinglecol \let\nofloat=\no@float\fi
}

\def\endtable{%
  \par\egroup 
  \egroup
  \setbox\TEMPBOX=\hbox to \TEMPDIMEN{%
    \eightpoint\nostb@ls{10pt}%
    \hss
    \vbox{%
      \hsize=\wd\ItemBOX
      \ifvoid\tab@cap@box
      \else
        \noindent\unhbox\tab@cap@box
        \vskip 5.5pt plus 6pt%
      \fi
      \box\ItemBOX
      \iftable@txt
        \vskip 10pt%
        \noindent\the\tab@txt@toks
        \global\table@txtfalse
      \fi
    }%
    \hss
  }%
  \ifsinglecol
    \if@here \midinsert\global\@herefalse\else \topinsert\fi
      \box\TEMPBOX
    \endinsert
  \else
    \global\setbox\ItemBOX=\box\TEMPBOX
    \JoinStack
    \Print{Processing source for table \the\ItemNUMBER}%
  \fi
}

\def\UnloadZoneA{%
\FirstZoneAtrue
 \Iteration=\z@
  \loop
   \ifnum\Iteration<\LengthOfStack
    \GetItemSTATUS{\Iteration}
    \ifnum\ItemSTATUS=\InZoneA
     \GetItemBOX{\Iteration}
     \ifFirstZoneA \vbox to \BodgeHeight{\vfil}%
     \FirstZoneAfalse\fi
     \unvbox\ItemBOX\ItemSep
     \LeaveStack{\Iteration}
     \else
     \advance\Iteration \@ne
   \fi
 \repeat
}

\def\UnloadZoneC{%
\Iteration=\z@
  \loop
   \ifnum\Iteration<\LengthOfStack
    \GetItemSTATUS{\Iteration}
    \ifnum\ItemSTATUS=\InZoneC
     \GetItemBOX{\Iteration}
     \ItemSep\unvbox\ItemBOX
     \LeaveStack{\Iteration}
     \else
     \advance\Iteration \@ne
   \fi
 \repeat
}


\def\ShowItem#1{
  {\GetItemAll{#1}
  \Print{\the#1:
  {TYPE=\ifnum\ItemTYPE=\Figure Figure\else Table\fi}
  {NUMBER=\the\ItemNUMBER}
  {SPAN=\ifnum\ItemSPAN=\Single Single\else Double\fi}
  {SIZE=\the\ItemSIZE}}}
}

\def\ShowStack{%
 \Print{}
 \Print{LengthOfStack = \the\LengthOfStack}
 \ifnum\LengthOfStack=\z@ \Print{Stack is empty}\fi
 \Iteration=\z@
 \loop
 \ifnum\Iteration<\LengthOfStack
  \ShowItem{\Iteration}
  \advance\Iteration \@ne
 \repeat
}

\def\B#1#2{%
\hbox{\vrule\kern-0.4pt\vbox to #2{%
\hrule width #1\vfill\hrule}\kern-0.4pt\vrule}
}


\newif\ifsinglecol   \singlecolfalse

\def\onecolumn{%
  \global\output={\singlecoloutput}%
  \global\hsize=\PageWidth
  \global\vsize=\PageHeight
  \global\ColumnWidth=\hsize
  \global\TextLeading=12pt
  \global\Leading=12
  \global\singlecoltrue
  \global\let\onecolumn=\relax
  \global\let\footnote=\sing@footnote
  \global\let\vfootnote=\sing@vfootnote
  \ninepoint 
  \message{(Single column)}%
}

\def\singlecoloutput{%
  \shipout\vbox{\PageHead\vbox to \PageHeight{\pagebody\vss}\PageFoot}%
  \advancepageno
  \ifplate@page
    \shipout\vbox{%
      \sp@pagetrue
      \def\sp@type{plate}%
      \global\plate@pagefalse
      \PageHead\vbox to \PageHeight{\unvbox\plt@box\vfil}\PageFoot%
    }%
    \message{[plate]}%
    \advancepageno
  \fi
  \ifnum\outputpenalty>-\@MM \else\dosupereject\fi%
}

\def\ItemSep{\vskip\ItemSepamount\relax}

\def\ItemSepbreak{\par\ifdim\lastskip<\ItemSepamount
  \removelastskip\penalty-200\ItemSep\fi%
}


\let\@@endinsert=\endinsert 

\def\endinsert{\egroup 
  \if@mid \dimen@\ht\z@ \advance\dimen@\dp\z@ \advance\dimen@12\p@
    \advance\dimen@\pagetotal \advance\dimen@-\pageshrink
    \ifdim\dimen@>\pagegoal\@midfalse\p@gefalse\fi\fi
  \if@mid \ItemSep\box\z@\ItemSepbreak
  \else\insert\topins{\penalty100 
    \splittopskip\z@skip
    \splitmaxdepth\maxdimen \floatingpenalty\z@
    \ifp@ge \dimen@\dp\z@
    \vbox to\vsize{\unvbox\z@\kern-\dimen@}
    \else \box\z@\nobreak\ItemSep\fi}\fi\endgroup%
}


\def\gobbleone#1{}
\def\gobbletwo#1#2{}
\let\footnote=\gobbletwo 
\let\vfootnote=\gobbleone

\def\sing@footnote#1{\let\@sf\empty 
  \ifhmode\edef\@sf{\spacefactor\the\spacefactor}\/\fi
  \hbox{$^{\hbox{\eightpoint #1}}$}\@sf\sing@vfootnote{#1}%
}

\def\sing@vfootnote#1{\insert\footins\bgroup\eightpoint\b@ls{9pt}%
  \interlinepenalty\interfootnotelinepenalty
  \splittopskip\ht\strutbox 
  \splitmaxdepth\dp\strutbox \floatingpenalty\@MM
  \leftskip\z@skip \rightskip\z@skip \spaceskip\z@skip \xspaceskip\z@skip
  \noindent $^{\scriptstyle\hbox{#1}}$\hskip 4pt%
    \footstrut\futurelet\next\fo@t%
}

\def\footnoterule{\kern-3\p@ \hrule height \z@ \kern 3\p@}

\skip\footins=19.5pt plus 12pt minus 1pt
\count\footins=1000
\dimen\footins=\maxdimen

\def\note#1#2{%
  \let\@sf=\empty \ifhmode\edef\@sf{\spacefactor\the\spacefactor}\/\fi
  #1\insert\footins\bgroup
    \eightpoint\b@ls{10pt}\rm
    \interlinepenalty\interfootnotelinepenalty
    \splitmaxdepth\dp\strutbox \floatingpenalty\@MM
    \leftskip\z@skip \rightskip\z@skip \spaceskip\z@skip \xspaceskip\z@skip
    \noindent\footstrut #1$\,$#2\strut\par
  \egroup
  \@sf\relax}


\def\landscape{%
  \global\TEMPDIMEN=\PageWidth
  \global\PageWidth=\PageHeight
  \global\PageHeight=\TEMPDIMEN
  \global\let\landscape=\relax
  \onecolumn
  \message{(landscape)}%
  \raggedbottom
}


\output{%
  \ifLeftCOL
    \global\setbox\LeftBOX=\vbox to \ZoneBSize{\box255\unvbox\ZoneBBOX
      \ifvoid\footins\else
        \vskip\skip\footins\unvbox\footins\fi
    }%
    \global\LeftCOLfalse
    \MakeRightCol
  \else
    \setbox\RightBOX=\vbox to \ZoneBSize{\box255\unvbox\ZoneBBOX
      \ifvoid\footins\else
        \vskip\skip\footins\unvbox\footins\fi
    }%
    \setbox\MidBOX=\hbox{\box\LeftBOX\hskip\ColumnGap\box\RightBOX}%
    \setbox\PageBOX=\vbox to \PageHeight{%
      \UnloadZoneA\box\MidBOX\UnloadZoneC}%
    \shipout\vbox{\PageHead\vbox to \PageHeight{\box\PageBOX\vss}\PageFoot}%
    \advancepageno
    \ifplate@page
      \shipout\vbox{%
        \sp@pagetrue
        \def\sp@type{plate}%
        \global\plate@pagefalse
        \PageHead\vbox to \PageHeight{\unvbox\plt@box\vfil}\PageFoot%
      }%
      \message{[plate]}%
      \advancepageno
    \fi
    \global\LeftCOLtrue
    \CleanStack
    \MakePage
  \fi
}


\Warn{\start@mess}

\newif\ifCUPmtplainloaded 
\ifprod@font
  \global\CUPmtplainloadedtrue
\fi


\catcode `\@=12 




\def\jcd{Christensen-Dalsgaard}

\begintopmatter
\title{A study of possible temporal and latitudinal variations in the
properties of the solar tachocline}
\author{Sarbani Basu$^{1}$ and H. M. Antia$^2$}

\affiliation{$^1$ Astronomy department, Yale University, P.O. Box 208101 New Haven
CT 06520-8101, U. S. A.}
\smallskip
\affiliation{$^2$ Tata Institute of Fundamental Research, Homi Bhabha Road,
Mumbai 400 005, India}
\acceptedline{Accepted \ . Received \ }

\abstract{Temporal variations of the structure and the rotation rate of the
solar tachocline region are studied using helioseismic data
from the  Global Oscillation Network Group (GONG) and the  Michelson
Doppler Imager (MDI) obtained during the period 1995--2000.
We do not find any significant temporal variation in the depth of the convection
zone, the position of the tachocline or the extent of overshoot
below the convection zone. No systematic variation in
any other properties of tachocline, like width, etc., 
is found either.
Possibility of periodic variations in these properties is also
investigated.
Time-averaged results show that the tachocline
is prolate with a variation by about $0.02R_\odot$ in its position.
The depth
of the convection zone or the extent of overshoot does not show any
significant variation with latitude.
}

\keywords {Sun: oscillations -- Sun: interior -- Sun: rotation }

\maketitle

\section{Introduction}
Helioseismic data allow us to probe the structure and
rotation rate of the solar interior
(Gough et al.~1996; Thompson et al.~1996; Schou et al.~1998).
With the accumulation of Global Oscillation Network Group (GONG) and
Michelson Doppler Imager (MDI) data over the last five years, it 
has also  become
possible to study temporal variations in the rotation rate and other
properties of the  solar interior.
It is generally believed that the solar dynamo operates in the region
just below the convection zone, which is also the region where the
tachocline is located (Kosovichev 1996; Basu 1997).
The tachocline is defined to be the region where
the rotation rate undergoes a transition from differential rotation in
the convection zone to almost uniform rotation in the radiative interior
(Spiegel \& Zahn 1992). As a result, one would expect temporal
variations associated with solar cycle to manifest in this region.
However, no definite changes have been detected in these layers
(Basu \& Antia 2000a).
Recently, Howe et al.~(2000) have reported a 1.3 year periodicity
in variation of equatorial rotation rate at  $r=0.72R_\odot$.
It is not clear if this period  is associated with solar
cycle related variations, or indeed why it should manifest only 
in a narrow latitude and radius range. Using similar data, 
Antia \& Basu (2000) did not find
any periodic or systematic changes in rotation rate in the tachocline
region. But both these studies are based on inversions of
data to obtain the  rotation
rate. This process is not very reliable in the tachocline region where the steep
gradient in the rotation rate  tends  to be smoothed by regularisation
applied in inversion techniques (Gough \& Thompson 1991; Antia, Basu
\& Chitre 1998).
The  properties of the solar tachocline have been
studied using forward modelling techniques (Kosovichev 1996; Basu 1997;
Antia et al.~1998; Charbonneau et al.~1999), which are probably
better suited to account for the steep variation in rotation rate.
Corbard et al.~(1998, 1999) have modified
inversion technique to account for sharp changes in the rotation rate.
Using a simple calibration technique to study the tachocline,
Basu \& Schou (2000) found that
the magnitude of the jump in the rotation rate across the tachocline increases
with solar activity. This result needs to be checked using other
techniques and with more data that are now available.

In this work, we attempt to use forward modelling techniques 
described by Antia et al.~(1998) to study whether different properties
of the tachocline vary with time.
We also look for possible periodic
changes  in the rotation rate around the tachocline.
In addition, we also study possible
temporal changes in solar structure in that region, in particular, changes in the
depth of the convection zone and the extent of overshoot below
the convection zone.
Since the depth of the convection zone can be measured
very accurately -- with statistical error of the order of $0.0001R_\odot$ --
it should be possible to detect even small variations in the 
structure of this region.

Apart from temporal variations in the properties of the tachocline, we also
attempt to find latitudinal variations in the position and the thickness of
the tachocline.  Antia et al.~(1998)
did not find any significant latitudinal variation in the position
or thickness of the tachocline. Although, the results did show
that the tachocline shifts upwards with latitude, this shift was
comparable to the error estimates. Charbonneau et al.~(1999)
found that the mean position of the tachocline moves upwards with latitude.
They found a shift of $(0.024\pm0.004)R_\odot$ in tachocline position
between the latitudes of $0^\circ$ and $60^\circ$, which is
comparable to the upper limit on its variation given by Antia et al.~(1998).
With the accumulation of data over 5 years it may be possible to
find out if this latitudinal variation is statistically significant.
Similarly, there are some indications that the thickness of tachocline
increases with latitude (Antia et al.~1998; Charbonneau et al.~1999) but
this variation is not statistically significant, at least, in the  results obtained
so far. Thus it is of interest to check if this variation
can be confirmed with longer data set.

If the position of the tachocline varies with latitude, then a natural question
is whether the depth of the convection zone also varies with
latitude. The relative position of the tachocline and the convection
zone base plays a crucial role in the theory of the tachocline.
Gough \& Kosovichev (1995) using data from Big Bear Solar
Observatory (Woodard \& Libbrecht 1993) have claimed a decrease by about
$0.02R_\odot$ in depth of the convection zone between the equator and
latitude of $60^\circ$. This is comparable to latitudinal variations
in the tachocline as found by Charbonneau et al.~(1999).
However, such a large variation would yield a strong
signal in the even-order splitting coefficients because of the resulting
asphericity.
Antia et al.~(2000a) did not find any significant signal in
asphericity around the
base of the convection zone, although they did not specifically look for
a signal from varying depth of convection zone.
Monteiro \& Thompson (1998) attempted to detect
latitudinal dependence in depth of convection zone and extent of
overshoot below the convection zone, but the results were not
conclusive.
Similarly, Basu (1997) and Antia et al.~(2000b) attempted to look for
any possible
magnetic field near the base of the convection zone, using the even
order splitting coefficients, but once again no signal was found in
the observed data. We therefore, try  to
check for any latitudinal variation in the depth of convection zone.

\section{The data}
We use data for GONG months 1--46
to determine the rotation rate
and the spherically symmetric structure in the solar
interior. Each of these data sets covers a period of 
108 days. We have used only the non-overlapping
sets of data for most of the work to ensure that each data set is independent.
Each GONG month covers a period of 36 days with month 1, starting
on 1995 May 7 and month 46 ending on 1999 November 17. There are 15
non-overlapping data sets covering this period, and these have 
been used  to study the temporal
variations in rotation rate and structure. However, for studying the oscillatory
changes in rotation rate we have also used all
data sets even though they overlap in time. There are  44 data sets centered 
on GONG months 2--45.
The data were obtained from the GONG Data Storage and Distribution
System.
In order to provide an
independent test of the results, we also use the data  from MDI.
Each of these 18 data sets was obtained from 72 non-overlapping days of
observations covering a period from 1996 May 1 to 2000 April 9
with some gaps corresponding to period when the Solar and Heliospheric
Observatory (SOHO), the space-craft on which MDI is located, was
not operational.

All these data sets (GONG and MDI) contain both the
mean frequencies and splitting coefficients for the observed p-modes.
The GONG data are described by Hill et al.~(1996), while the MDI data
are described by Schou (1999). 
The frequency
of an eigenmode of a given degree $\ell$, a given radial order  $n$, and
a given azimuthal order $m$ can be expressed in
terms of these splitting coefficients using the expansion 
$$
\nu_{n\ell m}
= \nu_{n\ell} + \sum_{j=1}^{j_{\rm max}} a_j (n,\ell) \,
{\cal P}_j^{(\ell)}(m),
\eqno\eqname\split
$$
where $\nu_{n\ell}$ is the mean frequency of the $(n,\ell)$ multiplet,
$a_j(n,\ell)$ are the splitting coefficients and 
${\cal P}_j^{(\ell)}(m)$ are orthogonal polynomials in $m$
(Ritzwoller \& Lavely 1991; Schou, \jcd\ \& Thompson 1994).
There is some ambiguity in the normalisation of the polynomials
${\cal P}_j^{(\ell)}(m)$. We have used the definition given
by Schou et al.~(1994).
The odd splitting coefficients, $a_1,a_3,\ldots$, are determined by
rotation rate in solar interior and have been used to infer the
rotation rate as a function of depth and latitude.
While the even splitting coefficients $a_2,a_4,\ldots$
are determined by  magnetic fields
and other aspherical perturbations in solar interior, as
well as by second order effects of rotation (Gough \& Thompson
1990).
The even splitting coefficients can be used to study the latitudinal
variations in the solar structure.
For the purpose of this work we have used only modes with frequencies
between 1500 and 3500 $\mu$Hz.

\section{The technique}

\subsection{Depth of the convection zone}
To detect possible changes in solar structure around the base of the
convection zone, we determine the depth of the convection zone using
the mean frequencies of the p-modes.
In lower part of the convection zone the temperature gradient equals
the adiabatic temperature gradient, while below the convection zone base
the temperature gradient is radiative.
The difference in the temperature gradient below and above the base of the
convection zone introduces a characteristic feature
in the sound speed difference between two models (or between a model
and the Sun) if they have different convection zone depths.
This signal can be used
to determine this depth and we use the method described by Basu
and Antia (1997) for this purpose.

We use the even-order splitting coefficients, which can be used to 
analyse departures from spherical symmetry (Gough 1993), to determine possible 
latitude-dependence of the depth of the convection
zone.
For simplicity, we only consider axisymmetric perturbations (with the
symmetry axis coinciding with the rotation axis) that are
symmetric about the equator. In this case, using the  variational
principle,  the difference in frequency between the Sun and a solar model for
a mode of a given order, degree and azimuthal order ($n$, $\ell$, and $m$)
can be written as:
$$\eqalign{
{\delta\nu_{nlm}\over\nu_{nlm}}&=
\int_0^R\;dr\;\int_0^{2\pi}\;d\phi\;\int_0^\pi \sin\theta\;d\theta\;\cr
&\left({\cal K}_{c^2,\rho}^{n\ell}(r){\delta c^2\over c^2}(r,\theta)+
{\cal K}_{\rho,c^2}^{n\ell}(r){\delta\rho\over\rho}(r,\theta)
\right)Y_\ell^m(Y_\ell^m)^*,\cr}
\eqno\eqname\eqinv
$$
where
$r$ is radius, $\theta$ is colatitude,
${\cal K}_{c^2,\rho}^{n\ell}(r)$ and ${\cal K}_{\rho,c^2}^{n\ell}(r)$
are the kernels for spherically symmetric perturbations
(Antia \& Basu 1994),
$Y_\ell^m$ are spherical harmonics denoting the angular dependence of
the eigenfunctions for spherically symmetric star.
The relative difference in squared sound speed, ${\delta c^2/ c^2}$ and
density, ${\delta\rho/\rho}$ between the Sun and a solar model
can be expanded in terms of even order Legendre polynomials:
$$\eqalign{
{\delta c^2\over c^2}(r,\theta)&=\sum_k c_k(r)P_{2k}(\cos\theta),\cr
{\delta \rho\over \rho}(r,\theta)&=\sum_k \rho_k(r)P_{2k}(\cos\theta),\cr}
\eqno\eqname\crhok
$$
where $c_k(r)$ and $\rho_k(r)$ are shorthand notations
for $({\delta c^2/ c^2})_k(r)$ and $({\delta\rho/\rho})_k(r)$, respectively.
The spherically symmetric component ($k=0$) causes frequency differences
that are independent of $m$ and thus only contribute to the mean frequency
of the $(n,\ell)$ multiplet. Higher order terms give frequencies that
are functions of $m$ and thus contribute to the splitting coefficients.

The angular integrals in Eq. \eqinv\ can be evaluated to give
$$
\eqalign{
&\int_0^{2\pi}\;d\phi\;\int_0^\pi \sin\theta\;d\theta\;Y_\ell^m(Y_\ell^m)^*
P_{2k}(\cos\theta)\cr
&\qquad\qquad ={1\over\ell}Q_{\ell k}{\cal P}_{2k}^{(\ell)}(m),\cr}
\eqno\eqname\qlk
$$
where $Q_{\ell k}$ depends only on $\ell,k$ and ${\cal P}_{2k}^{(\ell)}(m)$
are the orthogonal polynomials defined by Eq. \split.
The extra factor of $1/\ell$ ensures that $Q_{\ell k}$ approach a constant
value at large $\ell$.
Thus with this choice of expansion [Eq.~\crhok] the inversion problem
is decomposed
into independent inversions for each even splitting coefficient and
$c_k(r),\rho_k(r)$ can be computed by inverting the splitting
coefficient $a_{2k}$.
This would be similar to the 1.5d inversions used to determine  rotation rates
(Ritzwoller \& Lavely 1991),  called so  because
a two dimensional solution is obtained
as a series of one dimensional inversions:
$$
{\ell a_{2k}{(n,\ell)}\over\nu_{n\ell}Q_{\ell k}}=
\int_0^R{\cal K}_{c^2,\rho}^{n\ell}c_k(r)\;dr+
\int_0^R{\cal K}_{\rho,c^2}^{n\ell}\rho_k(r)\;dr.
\eqno\eqname\ainv
$$
To determine the sound speed or density at a particular colatitude $\theta$
we can combine these solutions using Eq.~\crhok.
Thus to invert for $\delta c^2/c^2$ at colatitude $\theta$, we can use the
usual equation for spherically symmetric case with the frequency
difference given by:
$$
\delta\nu=\delta\nu_{n\ell}+\sum_k{\ell a_{2k}(n,\ell)\over Q_{\ell k}}
P_{2k}(\cos\theta),
\eqno\eqname\delnu
$$
where $\delta\nu_{n\ell}$ is the difference in mean frequency for
the $n,\ell$ multiplet, between
the Sun and a solar model. Thus with this choice of $\delta\nu$ we
can apply the technique used by Basu \& Antia (1997) to determine
convection zone depth at a given colatitude.
 
\subsection{Overshoot below the convection zone}
In addition to
the depth of the convection zone, we attempt to determine changes
in the
extent of overshoot below the convection zone.
The sudden change in the temperature gradient at the base of the
convection zone introduces an oscillatory signal in the frequencies
as a function of radial order $n$ (Gough 1990).
The presence of an adiabatically stratified overshoot layer causes the 
temperature gradient to have a discontinuity, the magnitude of discontinuity
increasing with the extent of overshoot. Thus the amplitude of the
oscillatory signal also increases with the extent of overshoot.
To a first approximation, this signal has the form
$A\cos (2\tau\nu + \phi)$, where $\nu$ is the frequency of the mode,
$\tau$, the `frequency' of the signal is the acoustic depth of the
transition layer, $A$ is the amplitude and $\phi$  a phase.
This signal has been used earlier to estimate the
extent of overshoot below the solar convection zone
(Monteiro, \jcd\ \& Thompson 1994;
Basu, Antia \& Narasimha 1994). The oscillatory signal can be
amplified by taking the fourth
differences of the frequencies as a function of the radial order $n$
enabling a more precise measurement of the amplitude, $A$, of
oscillations (Basu et al.~1994).
Since the  amplitude of the signal  increases with increase in
extent of overshoot, it 
can be calibrated against amplitudes for models with known extents of
overshoot.  We use the method
described by Basu (1997) to isolate the oscillatory signal and measure
its characteristics.
Data sets used so far indicate that the amplitude of the
oscillatory signal is consistent with no overshoot, however,
this does not preclude any change of the amplitude with time.
The precision of   this measurement is much less than
that of the depth of the convection zone, but since there is some
suggestion that this amplitude may be varying with time (Monteiro et al.~1998)
it needs to be checked independently.

We can also calculate the latitude dependence of extent of overshoot
by adding the contribution from even splitting coefficients as given
in Eq.~\delnu\ to the frequencies. The resulting frequencies can
be used to calculate the fourth difference and the oscillatory
part can be isolated as explained above to calculate the amplitude.
Any variation in amplitude of this oscillatory signal can be
attributed to change in extent of overshoot with latitude.
Since the errors in estimating the extent of overshoot are much
larger than those in estimating the depth of the convection zone,
probability of detecting either temporal or latitudinal variation
in extent of overshoot is much smaller than those in convection zone
depth. But for completeness we have attempted to study this
variation too.

\subsection{The tachocline}
To determine the properties of tachocline we use the three techniques
described by Antia et al.~(1998), which are (1) a calibration method
in which the  properties at each latitude are determined by direct 
comparison with models;
(2) a  one dimensional (henceforth, 1d) annealing technique in which  the
parameters defining the tachocline at each latitude  are determined by
a nonlinear least squares minimization using simulated annealing method
and (3) a  two-dimensional (henceforth, 2d) annealing technique, where the entire 
latitude dependence of tachocline
is fitted simultaneously, again using simulated annealing.
The properties we are interested in are 
the position and the thickness of the tachocline and 
the change in rotation rate  across the tachocline. We study these
properties  as a function of latitude using all the techniques listed
above.
The use of three different techniques allows us a check on the results.
In all techniques the tachocline is represented
by a model of the form (cf., Antia et al.~1998),
$$
\Omega_{\rm tac}=
{\delta\Omega\over {1+\exp[(r_t-r)/w]}},\eqno\eqname\thi$$
where
$\delta\Omega$ is the jump in the rotation rate across the tachocline,
$w$ is the half-width of the transition layer, and $r_t$ the
radial distance of the mid-point of the transition region. 
A positive value of $\delta\Omega$ implies that the rotation rate in the
convection zone is larger than that in the radiative zone.
The quantities $\delta\Omega$,
$w$ and $r_t$ could be a function of latitude. It should be noted that the
model for tachocline used by us is different from that used by
Kosovichev~(1996) or Charbonneau et al.~(1999) and in particular,
the definition of width in the two models is significantly different
(see Antia et al.~1998).  For the annealing
techniques a model for smooth part of the rotation rate is also
required and we use the same model as that used by Antia et al.~(1998):
$$\Omega(r)=\cases{\Omega_c+B(r-0.7)& if $r\le 0.95$\cr
\Omega_c+0.25B-C(r-0.95)& if $r>0.95$.\cr}
\eqno\eqname\tachsm
$$
Here $r$ is the radius in units of solar radius,
$\Omega_c$ is the rotation rate in radiative interior,
$B$ is the average gradient of the rotation rate in lower part of 
the convection zone and
$C$ is the average gradient in the near surface shear layer. 

\beginfigure{1}
\centerline{\epsfysize=8.00 true cm\epsfbox{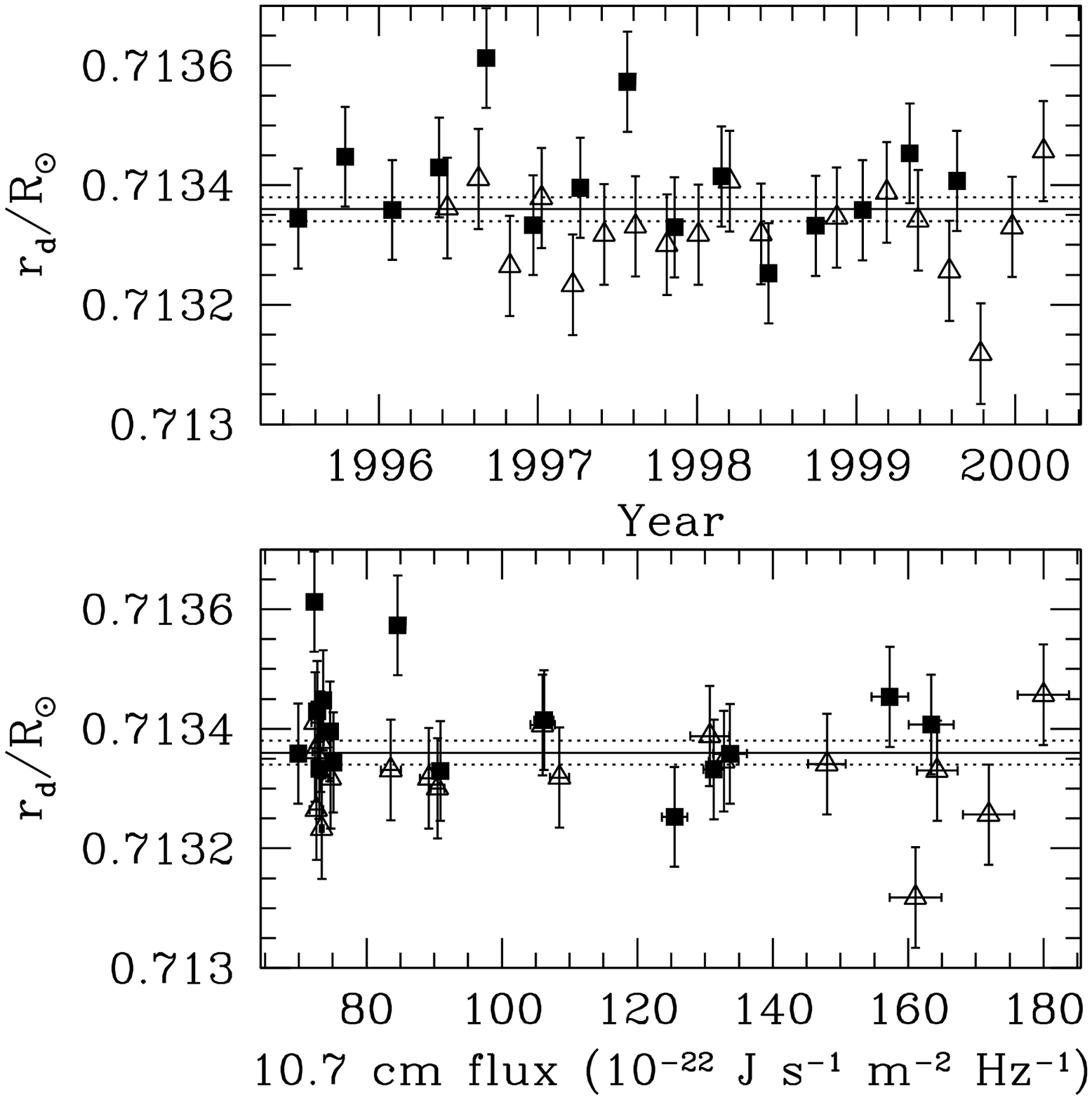}}
\caption{{\bf Figure 1.} The position of the base of the solar
convection zone plotted as a function of time and the 10.7 cm radio
flux, which is an indicator of solar activity. The squares represent GONG
and triangles denote MDI  data. The continuous, horizontal line is
the average of all the results, with the $1\sigma$ limits marked
by dotted lines.
}
\endfigure

\section{Results}
\subsection{Depth of the convection zone}

We calculate the position of the base of the convection zone ($r_d$)
using the frequencies from each data set from GONG and MDI and the
results are shown in Fig.~1. 
The error-bars shown in Fig.~1 represent uncertainties due to those
in the data, i.e.,
statistical errors. There are systematic errors which can be up to a
factor of five larger than the statistical errors (Basu 1998).
The systematic error for each data set should be the same
and hence does not affect any conclusion about variations with
time.
There is very good agreement between results obtained using GONG and
MDI data. There does not appear to be any systematic
pattern in these temporal variations. There are some oscillations in these
values as a function of time. In order to check for possible periodicities
we take the Fourier transform of $r_d(t)$ and we do not see any
significant peak in the resulting power spectrum at any frequency.
Thus the variations do not appear to be periodic. The temporal average
over all the 33 data sets from GONG and MDI gives 
$r_d=(0.71336\pm0.00002)R_\odot$, which agrees well with the
earlier estimates of \jcd, Gough \& Thompson~(1991), Basu \& Antia (1997)
and Basu (1998).

The lower panel in Fig.~1 shows the
results plotted against the mean 10.7 cm Radio flux during the time
interval covered by each data set as obtained from the
Solar Geophysical data web page (www.ngdc.noaa.gov/stp/stp.html) of the
US National Geophysical Data Center. The 10.7 cm flux from the
Sun is known to track solar activity.
There does not appear to be any correlation between the position
of convection zone base and the 10.7 cm flux.
In fact, the  correlation coefficient between these quantities
is only $-0.18$. 
All these results suggest that  there is no  systematic
variation in the depth of the convection zone with solar
activity.
An  upper limit on any
possible variation would be comparable to the error estimates in
individual measurements, which is $0.0001R_\odot$ or 70 km.
To get a better limit on any possible variation in the depth of the
CZ with solar activity, we divided the data into two groups,
those with 10.7 cm flux lower than 100 flux units (the unit being 
$10^{-22}$ J s$^{-1}$ m$^{-2}$ Hz$^{-1}$) and those with
10.7 cm flux of above 120 flux units. Nine data sets each from
GONG and MDI fall into the first category, and seven MDI sets
and five GONG sets fall into the second category. The low
activity data sets indicate that the CZ base is at
$r_d=(0.71337\pm0.00002)R_\odot$, the higher activity set gives
$r_d=(0.71333\pm0.00003)R_\odot$. Thus the variation in convection
zone depth is about $(28\pm 25)$ km, between these sets with
mean 10.7 cm flux of $77.2\pm7.1$ and $150.0\pm18.7$ flux units,
respectively.

As explained in Section~3.1, we can calculate the depth of the convection
zone at different latitudes using the even order splitting coefficients.
We have calculated $r_d$ at several different latitudes for all the
data sets and these results
also do not show any significant temporal variation at any latitude.
In this case, the error estimates are larger and hence the corresponding
upper limit on possible temporal variations would also be larger.

\beginfigure{2}
\centerline{\epsfysize=8.00 true cm\epsfbox{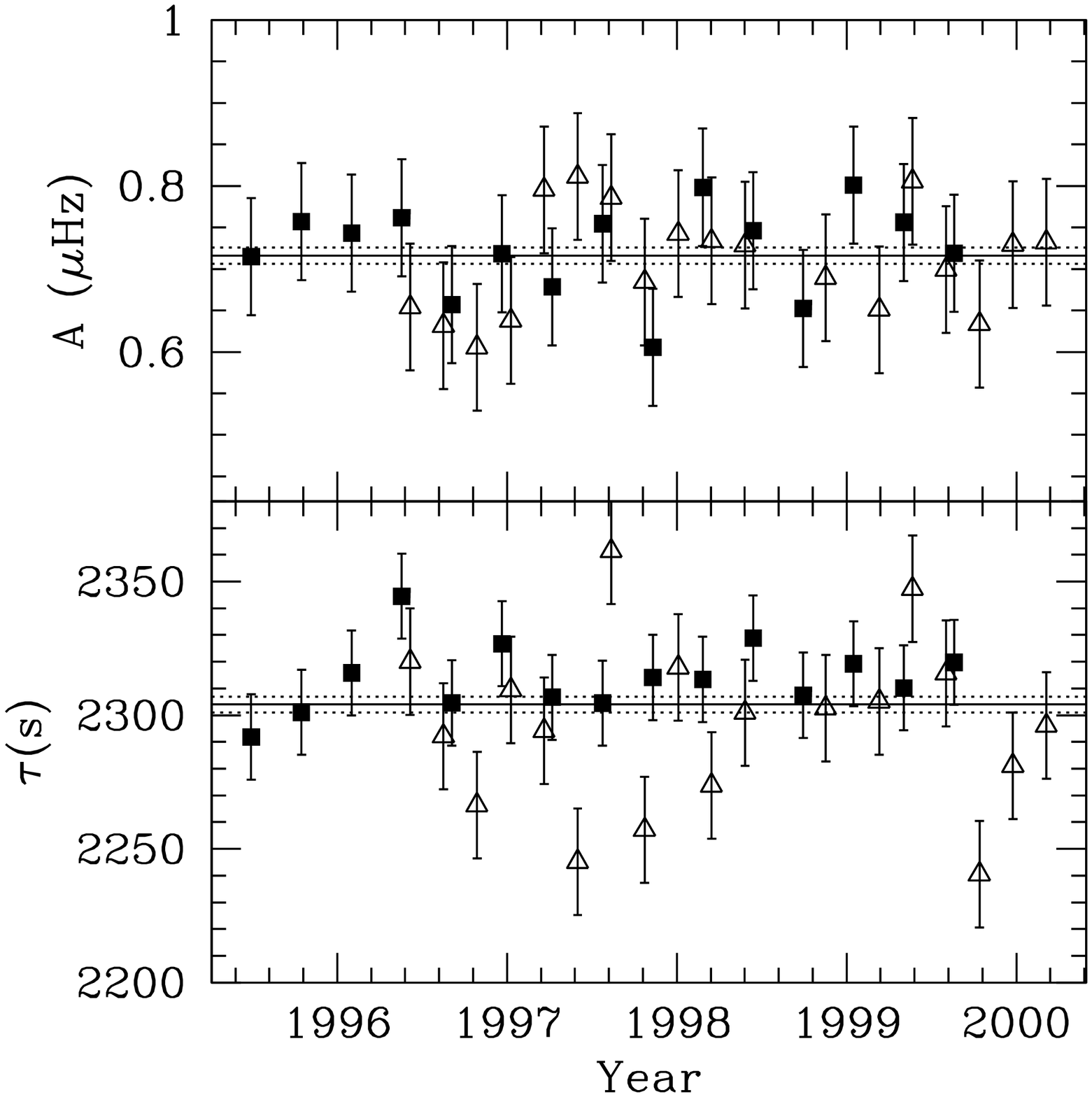}}
\caption{{\bf Figure 2.} The amplitude, $A$ and the frequency, $\tau$ of the 
oscillatory signal in the fourth difference of frequencies plotted as
a function of time for GONG (squares) and MDI (triangles) data.
The continuous, horizontal line is
the average of all the results, with the $1\sigma$ limits marked
by dotted lines.
}
\endfigure

\beginfigure*{3}
\centerline{\epsfysize=15.00 true cm\epsfbox{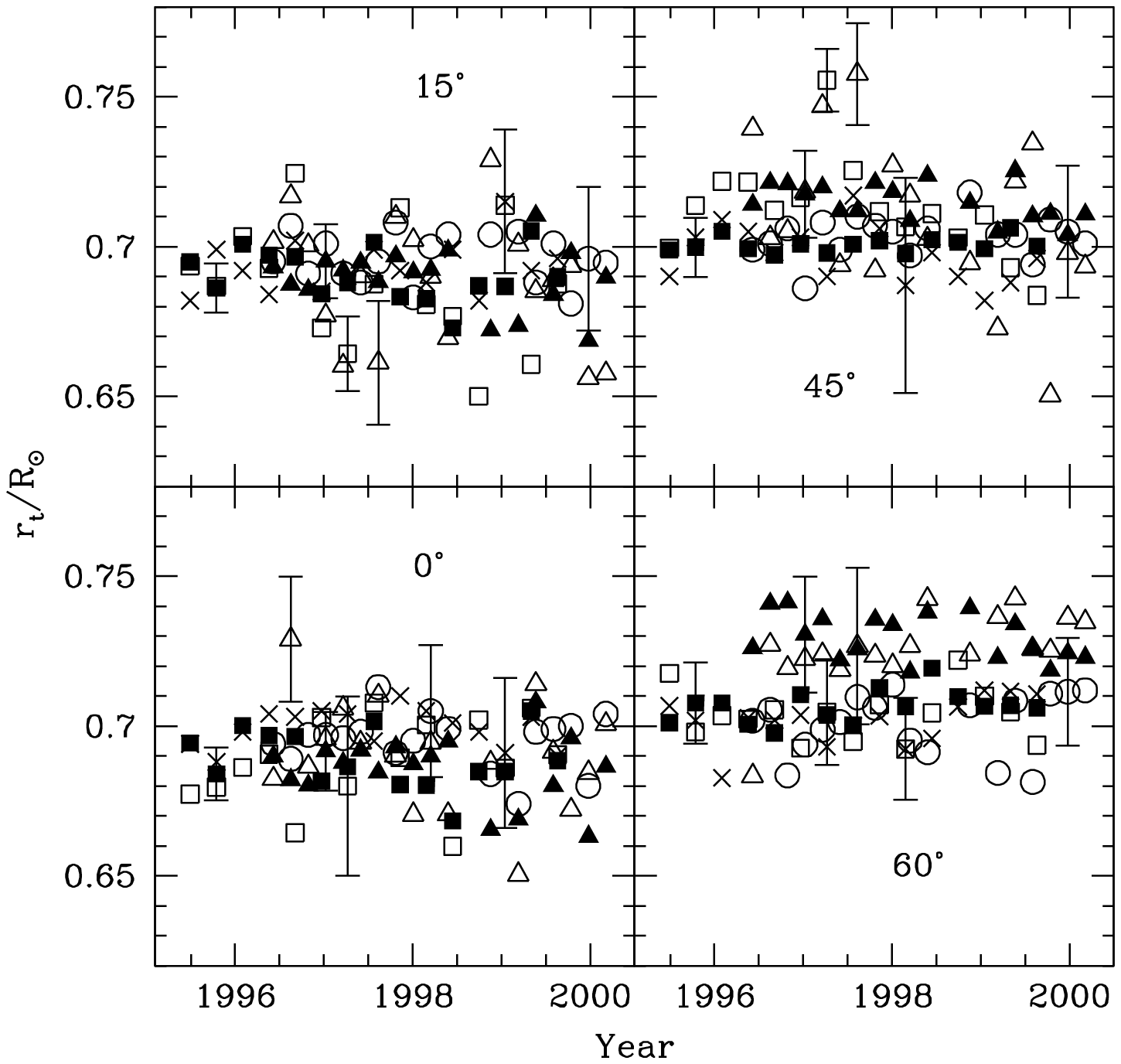}}
\caption{{\bf Figure 3.} The mean radial position of the tachocline
at a few selected latitudes. The
crosses and circles show the results from calibration method
for GONG and MDI data, the open squares and triangles
show the 1d annealing results from GONG and MDI data, while the
filled squares and triangles show the results
from 2d annealing for GONG and MDI data,
For clarity only one representative error-bar
is shown for each technique.
}
\endfigure

\beginfigure*{4}
\centerline{\epsfysize=15.00 true cm\epsfbox{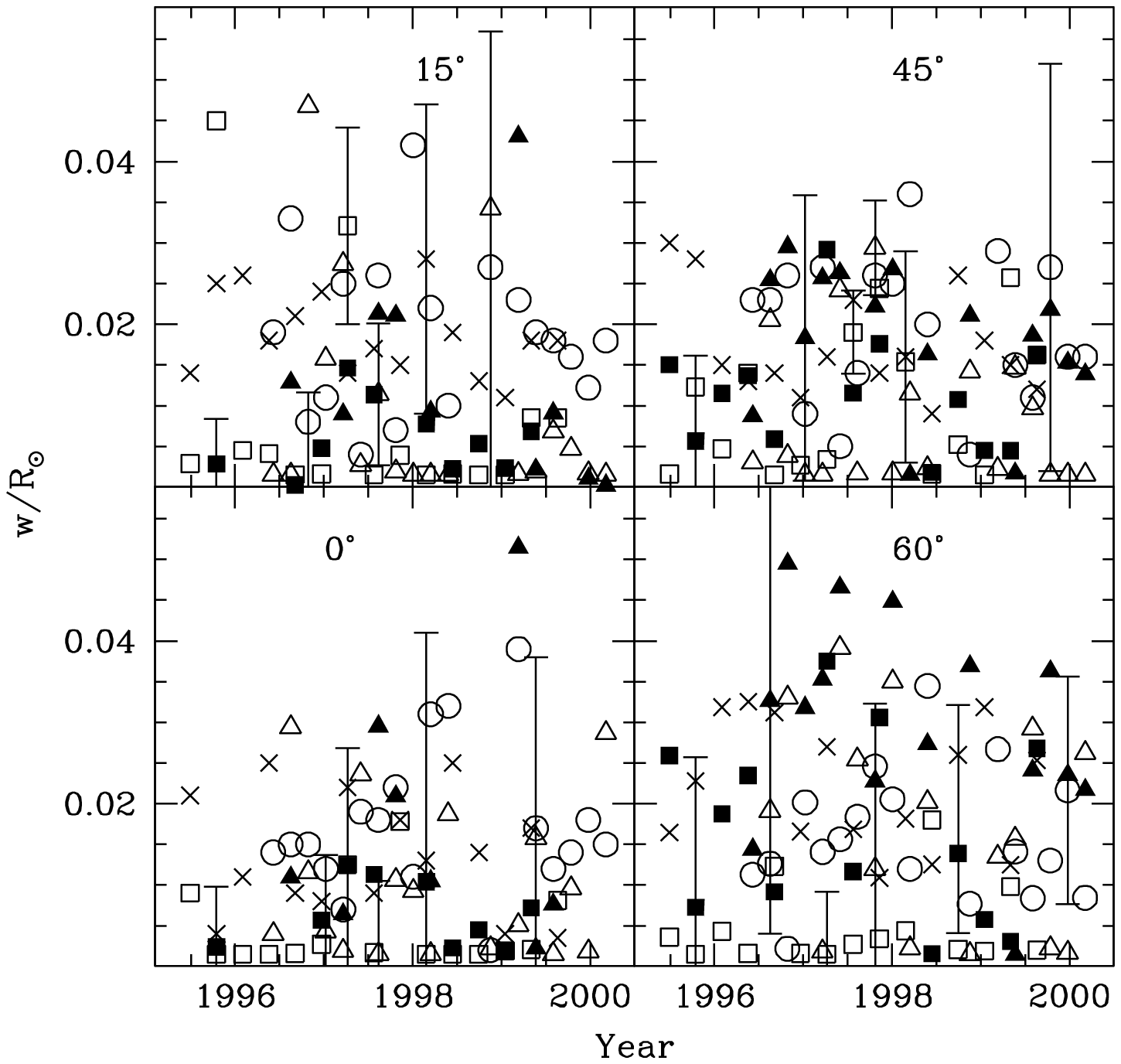}}
\caption{{\bf Figure 4.} The width of the tachocline 
at a few selected latitudes. 
The different styles of the points have the same meaning as in Fig.~3.
}
\endfigure

\beginfigure*{5}
\centerline{\epsfysize=15.00 true cm\epsfbox{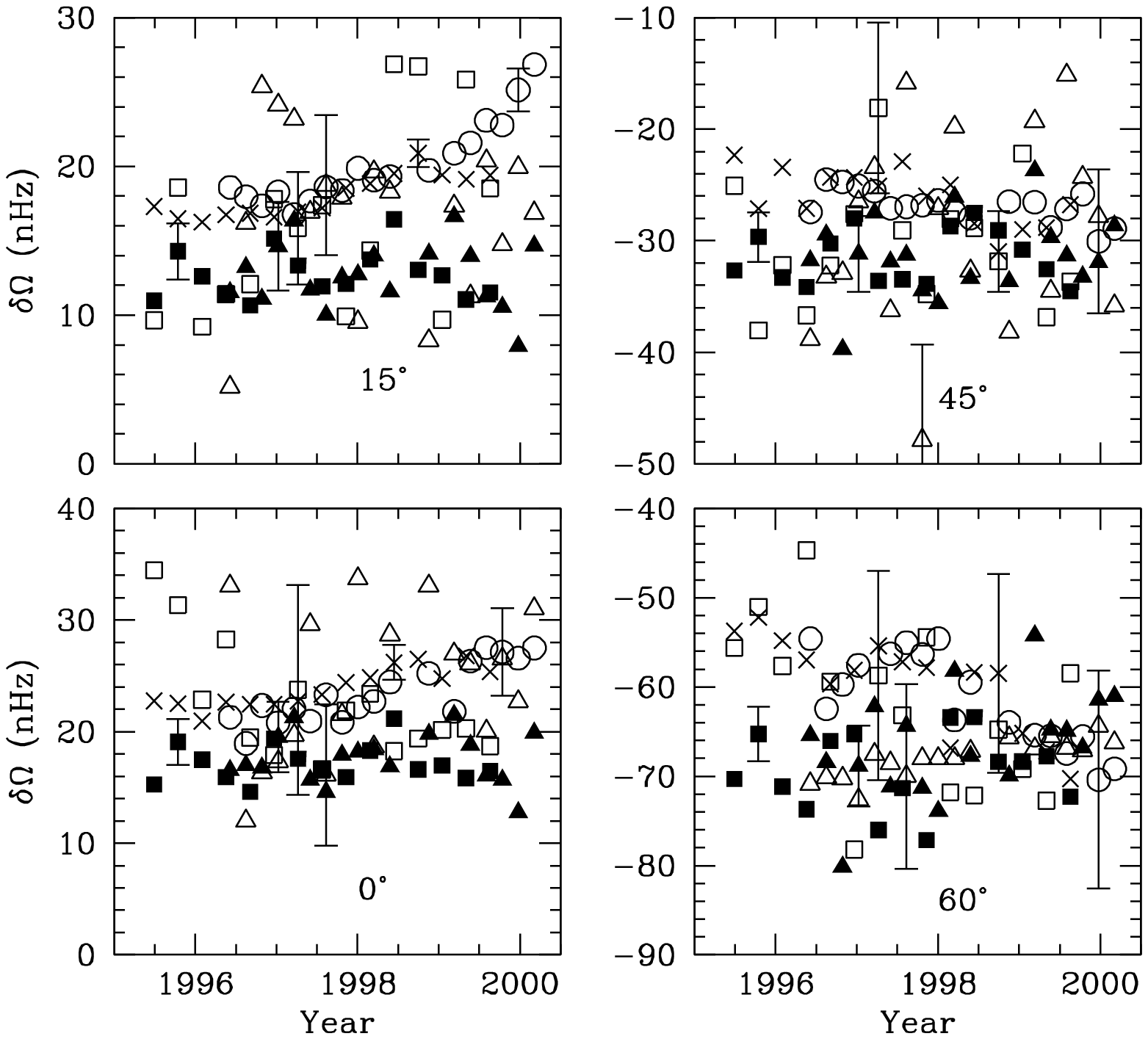}}
\caption{{\bf Figure 5.} The jump in rotation rate across the tachocline 
at a few selected latitudes. 
The different styles of the points have the same meaning as in Fig.~3.
}
\endfigure

\subsection{Overshoot below the convection zone}
We have used all independent data sets available to us to check whether there is
any temporal variation in the extent of overshoot
below the solar convection zone. For this purpose we use frequencies
for $p$-modes with $5\le\ell\le20$ and $2\le\nu\le3.5$ mHz.
Figure~2 shows the mean amplitude, $A$ of the
oscillatory part in the fourth difference of the frequencies due to the
transition at the base of the overshoot layer, as a function of time.
This amplitude would be related to the extent of overshoot below the
convection zone. There is good agreement between the results obtained
using GONG and MDI data sets.  Once again there is no clear trend
and the oscillatory features are not periodic.
The time-average of the amplitude as obtained from averaging over result
from all sets 
is $0.716\pm0.010$ $\mu$Hz, which is in reasonable agreement with the
earlier results obtained by Basu et al.~(1994) and Basu (1997).
This amplitude is significantly less than that in a solar model
without overshoot below the convection zone ($A=0.805$ $\mu$Hz).
The frequencies
in both GONG and MDI data have been calculated by assuming that
the peak profiles in power spectra are symmetric. However, it is
well known that the peaks are actually asymmetric and use of
symmetric peak profiles causes frequencies to be shifted away from
true value. Basu and Antia~(2000b) found that the use of symmetric
profile tends to underestimate the amplitude of the oscillatory signal.
This may explain the apparent contradiction that the amplitude for observed data
is significantly smaller than that in a solar model without overshoot.
If we assume that the use of asymmetric profile increases the amplitude
by the same amount as that found by Basu and Antia~(2000b) then the
resulting amplitude will be consistent with that in a solar model
without overshoot.

Figure~2 also shows how the frequency, $\tau$ of the oscillatory part in frequencies
due to the transition at the base of the overshoot layer changes
as a function of time.
This frequency is expected to be the acoustic depth of the base of
overshoot layer. Once again we do not see any systematic variation
in the acoustic depth with time.
The average of the results obtained using  all  33 data sets 
is $\tau=2304\pm3$ s, which is in reasonable agreement with the
earlier results obtained by Basu et al.~(1994) and Basu (1997).

As explained in Section~3.2 we can also calculate the extent of overshoot
at different latitudes, using the information from even splitting
coefficients. But in this case the errors are too large and it is
difficult to isolate the small oscillatory signal and hence no
definitive results could be obtained. 
However, we may not expect any time-variation in the overshoot results
since  the
depth of the convection zone, which can be determined much more
accurately, also doesn't show any temporal variation of either the
spherically symmetric value or values at individual latitudes.

\subsection{Temporal variations in the tachocline}
Having failed to find any systematic temporal variations in the spherically
symmetric structure near the base of the convection zone, we attempt
to look for variations in the rotation rate that define the tachocline.
Using each data set from GONG and MDI we determine the position,
width and jump across the tachocline using all the three techniques
mentioned in Section~3.3. For this purpose we use only those modes which
have lower turning points in the range 0.6--0.9$R_\odot$.
The results at a few selected latitudes are
shown in Figs.~3--5. There is a general agreement between different
techniques and data sets, as all the results are within respective
error estimates. However, some systematic differences between different
data sets and techniques can be seen, particularly at high latitudes.
These differences are comparable to error estimates in individual
values but if temporal mean is taken over all values the differences
could be significant. The origin of these systematic differences is
not clear.
There is no systematic temporal variation in either the
position or the width of the tachocline in any of the results.
From Fig.~3, it is clear that there is some latitudinal variation in
the position of the tachocline as the mean position moves upwards with
latitude. We will discuss this latitudinal variation
in the next subsection.
The jump in the rotation rate across the tachocline 
obtained by  the calibration method appears
to show a steady increase with time at low latitudes,
as was seen by Basu \& Schou (2000) using the same technique. The,
increase however,  is comparable to error
limits and is not seen in results obtained from annealing technique.
Hence the significance of this trend is not clear.
This trend is not very clear even in results obtained using calibration
method at latitudes of $45^\circ$ and $60^\circ$.

\beginfigure{6}
\centerline{\epsfysize=8.0 true cm\epsfbox{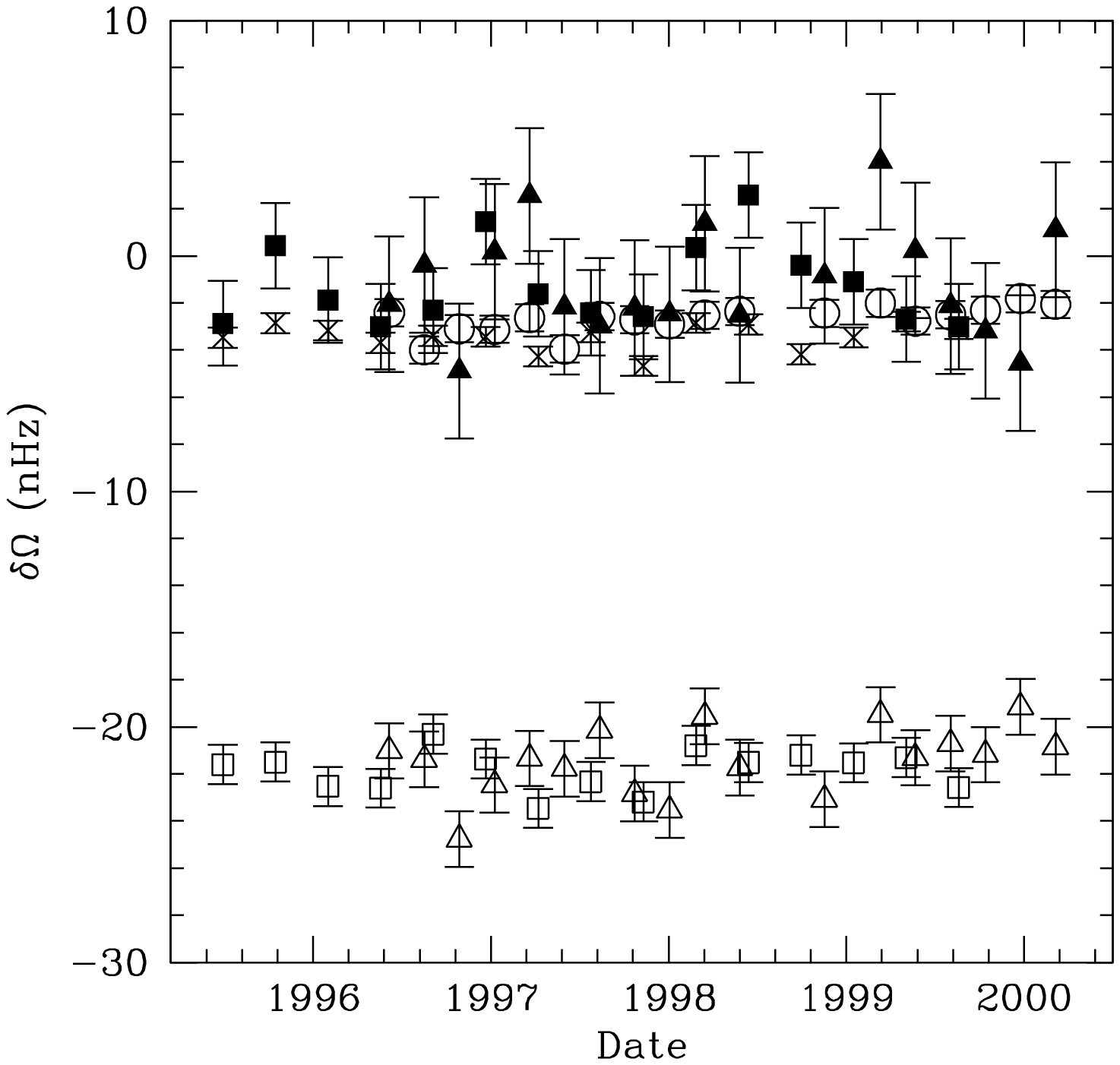}}
\caption{{\bf Figure 6.} The three components of the jump in rotation
rate across the tachocline as obtained using 2d annealing technique.
The filled squares and triangles show $\delta\Omega_1$ from GONG
and MDI data respectively. The open squares and triangles show
$\delta\Omega_3$ from GONG and MDI data respectively, while the
crosses and circles show $\delta\Omega_5$ from GONG and MDI data
respectively.
}
\endfigure

The values of the jump in the rotation rate across the tachocline
obtained using the 2d annealing technique 
appear to show an oscillatory behaviour at low latitudes.
This can be traced back  to oscillations in the latitudinally independent
component of jump. Following Antia et al.~(1998),  $\delta \Omega$
in the 2d annealing technique can be expanded  as
$$
\delta\Omega(\theta)=\delta\Omega_1+\delta\Omega_3P_3(\theta)
+\delta\Omega_5P_5(\theta),
\eqno\eqname\jump
$$
where
$$\eqalign{
P_3(\theta)&=5\cos^2\theta-1,\cr
P_5(\theta)&=21\cos^4\theta-14\cos^2\theta+1.\cr}
\eqno\eqname\pn
$$
Fig.~6 shows the three components obtained from GONG and MDI data
as a function of time. There are some oscillations in the first
component $\delta\Omega_1$ which appear to be similar to that
seen by Howe et al.~(2000) in the equatorial rotation rate at 
$r=0.72R_\odot$. It may be noted that we see these oscillations in
the spherically symmetric component which is independent of latitude
and its phase is also different from that found by Howe et al.~(2000).
In order to check the significance of these oscillations we need to
look at other parameters fitted in 2d annealing method. It turns out
that $\delta\Omega_1$ is anti-correlated with $\Omega_c$,  the
rotation rate in the radiative zone below the tachocline.
This anti-correlation is due to the fact that it is difficult to
distinguish between the effects of these two parameters in the
tachocline model that is used for fitting.
This behaviour is caused by the fact that the jump in the spherically
symmetric component of the rotation rate across the tachocline is very small
and as a result, it is difficult to distinguish
between the  sharp gradient of the rotation rate  in the tachocline region and 
the  smoother gradient in the convection zone.
Figure~7 shows $\Omega_c$
as well as the sum $\Omega_c+\delta\Omega_1$, which is  the
spherically symmetric component of rotation rate just above the
tachocline. It can be seen that the variations in $\Omega_c+\delta\Omega_1$
are much less than those in either   $\Omega_c$ or $\delta\Omega_1$.

\beginfigure{7}
\centerline{\epsfysize=9.50 true cm\epsfbox{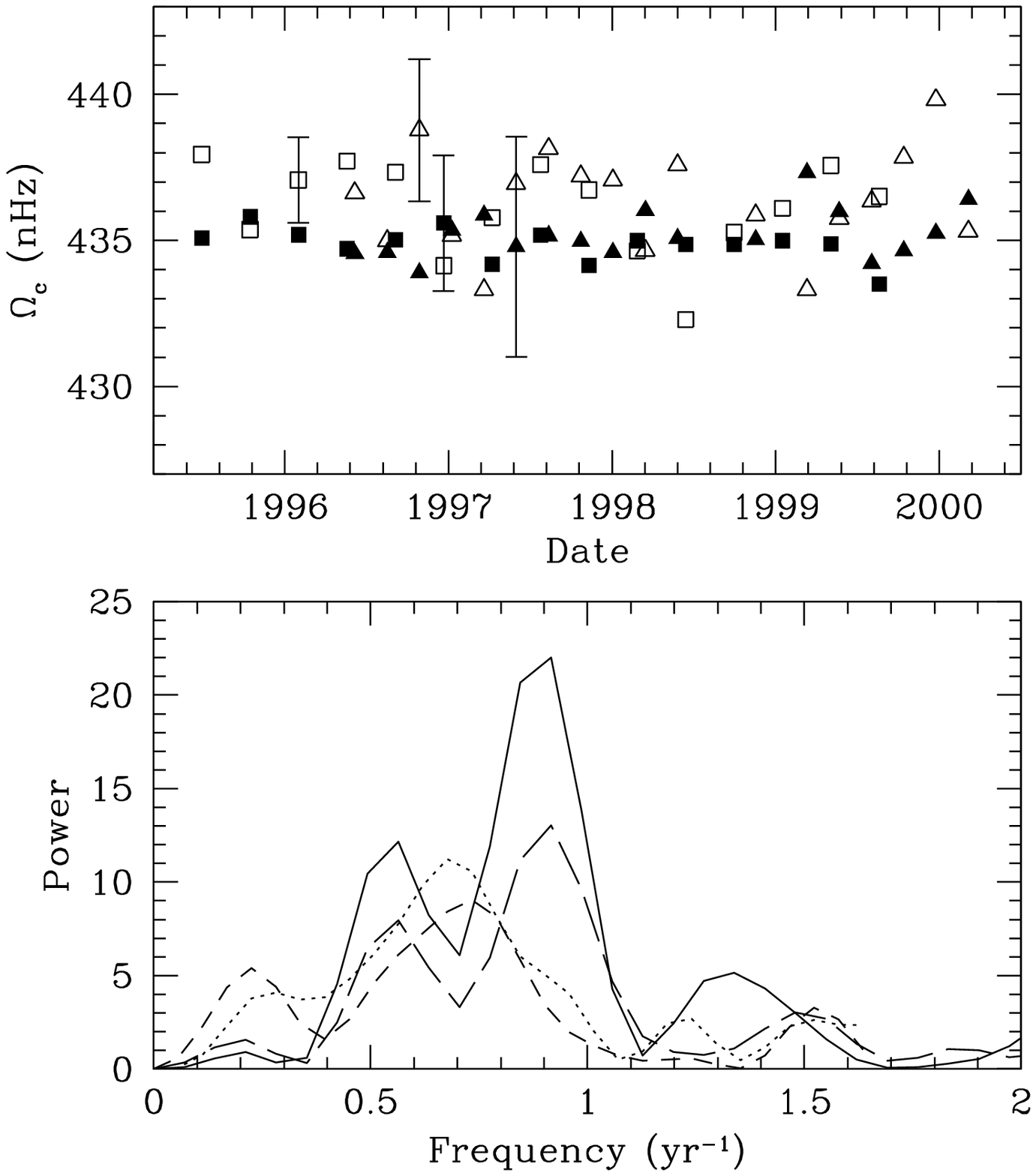}}
\caption{{\bf Figure 7.} The upper panel shows the
latitudinally averaged rotation rate in
radiative interior, $\Omega_c$ and that just above the
tachocline ($\Omega_c+\delta\Omega_1$) as obtained from the 2d
annealing technique. The open squares and triangles represent $\Omega_c$
obtained using the GONG and MDI data respectively, while filled symbols
show $\Omega_c+\delta\Omega_1$.
The lower panel shows the Fourier transform of $\Omega_c$ and
$\delta\Omega_1$.
The continuous and dotted lines show the power in $\delta\Omega_1$ from
MDI and GONG data respectively, while the long-dashed and short-dashed
lines show the power in $\Omega_c$ from MDI and GONG data.
Error estimates on these power spectra are not shown as the estimated
error is larger than the peak heights.
}
\endfigure

\beginfigure{8}
\centerline{\epsfysize=9.00 true cm\epsfbox{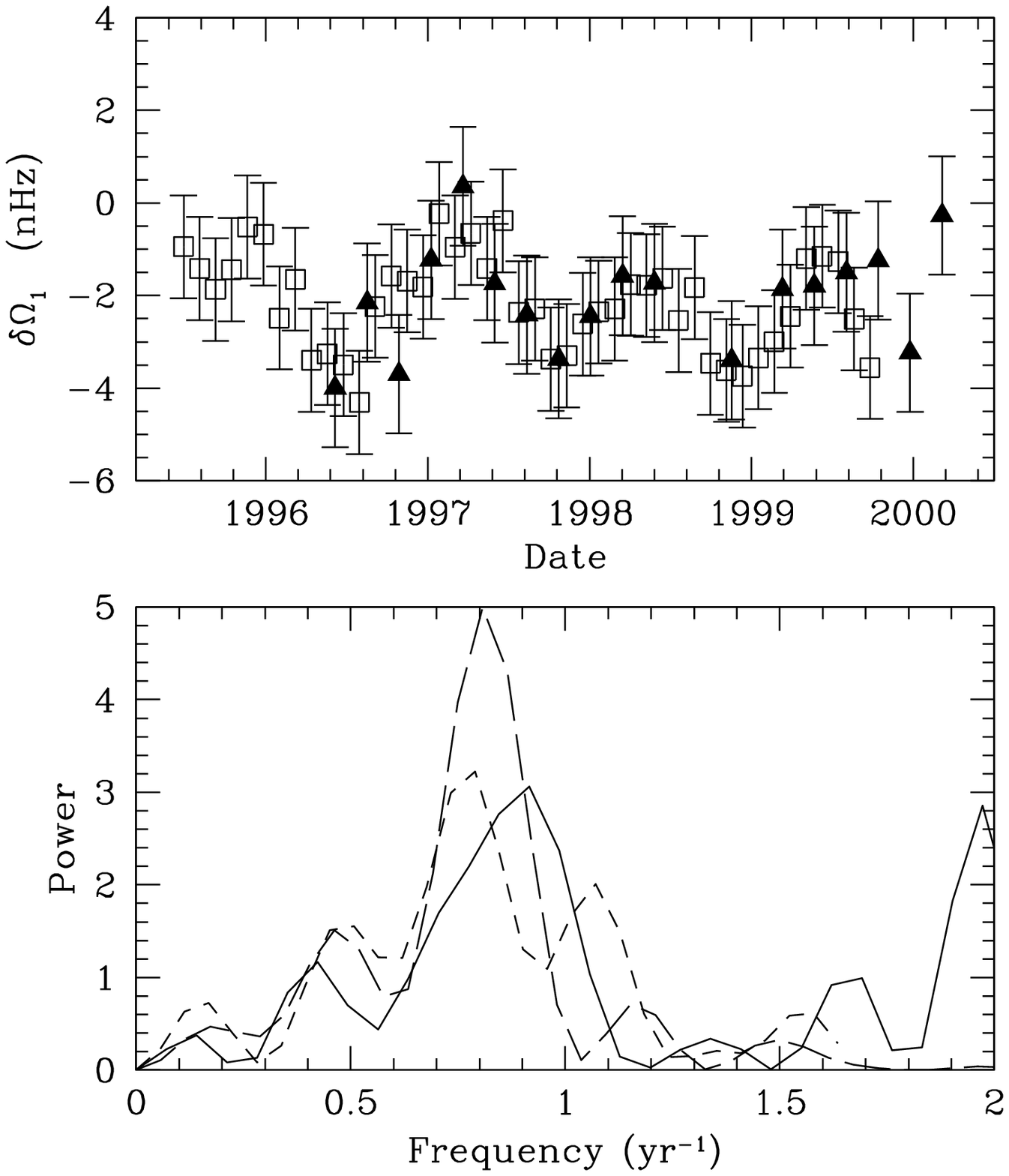}}
\caption{{\bf Figure 8.} The upper panel shows the jump in
latitudinally independent
component of rotation rate across the tachocline, $\delta\Omega_1$
obtained by fitting only the splitting coefficient $a_1$.
The squares and triangles, respectively represent the values obtained from
GONG and MDI data. The lower panel shows the Fourier transform of 
$\delta\Omega_1$ from MDI (continuous line) and GONG (short-dashed line) data
using non-overlapping data sets, while the long-dashed line shows the
spectra using all data sets from GONG.
}
\endfigure

In order to test if the oscillatory variation in $\delta\Omega_1$
or $\Omega_c$ are periodic, we take the Fourier transform of these
results (after subtracting out the temporal mean)
and the resulting power spectra are shown in the lower panel of Fig.~7.
It is clear
that the spectra of $\delta\Omega_1$ and $\Omega_c$
from GONG data shows a
moderately clear peak around a frequency of 0.7 ${\rm yr}^{-1}$ or a
period of 1.4 years. This is comparable to the period found by
Howe et al.~(2000) but the peak is rather broad and not very significant
statistically.
The $1\sigma$ error estimates on these peaks turns out to be larger than
the peak height and hence these peaks are not likely to be significant.
Moreover, the peak is not clearly present in MDI data. The Fourier
spectrum of MDI results shows two peaks
at frequencies of $0.6$ and 0.9 ${\rm yr}^{-1}$, with the second peak
being more dominant. 
The combination $\Omega_c+\delta\Omega_1$ does not show any clear peak
for either GONG or MDI data. Hence, this spectrum is not shown in the
figure.

Since the latitudinally independent component of rotation rate should 
contribute only to the splitting coefficient $a_1$, we also try to
fit $a_1$ separately to check if the oscillations are present.
Furthermore, following Howe et al.~(2000) we fit all GONG data sets centered at
GONG months 2--45, including those that overlap in time.
Since the jump in $a_1$ is rather small it is not possible to fit
all the parameters used to define the tachocline using  $a_1$ alone.  We,
therefore, keep
the position and half-width of tachocline fixed at representative values
of $0.69R_\odot$ and $0.009R_\odot$ to obtain other parameters
including $\delta\Omega_1$ and $\Omega_c$.
These results are shown in Fig.~8.
We have verified that these
results are not sensitive to the choice of position and thickness.
The oscillations in $\delta\Omega_1$ are still present in these
results, though it is not clear if these are strictly periodic. 
There is very good agreement between GONG and MDI data points at
all times and these oscillations are similar to that found by
Howe et al.~(2000) in equatorial rotation rate at $r=0.72R_\odot$.
The phase and amplitude of these oscillations are also similar to that
found by Howe et al.~(2000).
To check for periodicities we take the Fourier transform (after
subtracting out the temporal mean) and the
results are also shown in Fig.~8. This figure also shows the Fourier
transform of results obtained from  only non-overlapping sets of GONG data. 
It can be seen
that in both cases we get similar peaks in the power spectra, thus it
may not be essential to use all data sets in order to identify periodicity.
However, the significance of the peaks is, of course, different
when all data sets are used. 
The GONG and MDI spectra
show a peak at frequencies around $0.8$ and $0.9$ yr$^{-1}$,
respectively. The peak heights are smaller than $1\sigma$ error
estimates when only non-overlapping data sets are used for GONG.
The peak in the Fourier spectrum of  results obtained from  all GONG data sets,
is about $1.5\sigma$, but
its significance is not clear since the errors are estimated by
assuming that all data sets are independent, which is not the case
because of the temporal overlap. Thus in all cases the significance of
the peak is questionable.
It may be noted that since we have data spanning about
4 years, the frequency resolution in the power spectra 
is around 0.25 yr$^{-1}$. Hence, to within this frequency resolution
the peaks in GONG and MDI spectra are close to the period of 1 year,
which may arise from annual variations in the data. Both GONG and MDI
data could be affected by the orbital motion of the Earth and one
may expect a periodicity of 1 year in the data. 
Furthermore, it may be noted that the peak height has
reduced by a factor of more than 3 between Figs.~7 and 8. This
is approximately the square of ratio of error estimates in the two
cases. This probably implies that at least, in Fig.~7, the peak is
not significant.
In the case of fit to only $a_1$ also,  $\Omega_c$ and $\delta\Omega_1$
are anti-correlated and their Fourier spectra are similar.

\beginfigure{9}
\centerline{\epsfysize=10.00 true cm\epsfbox{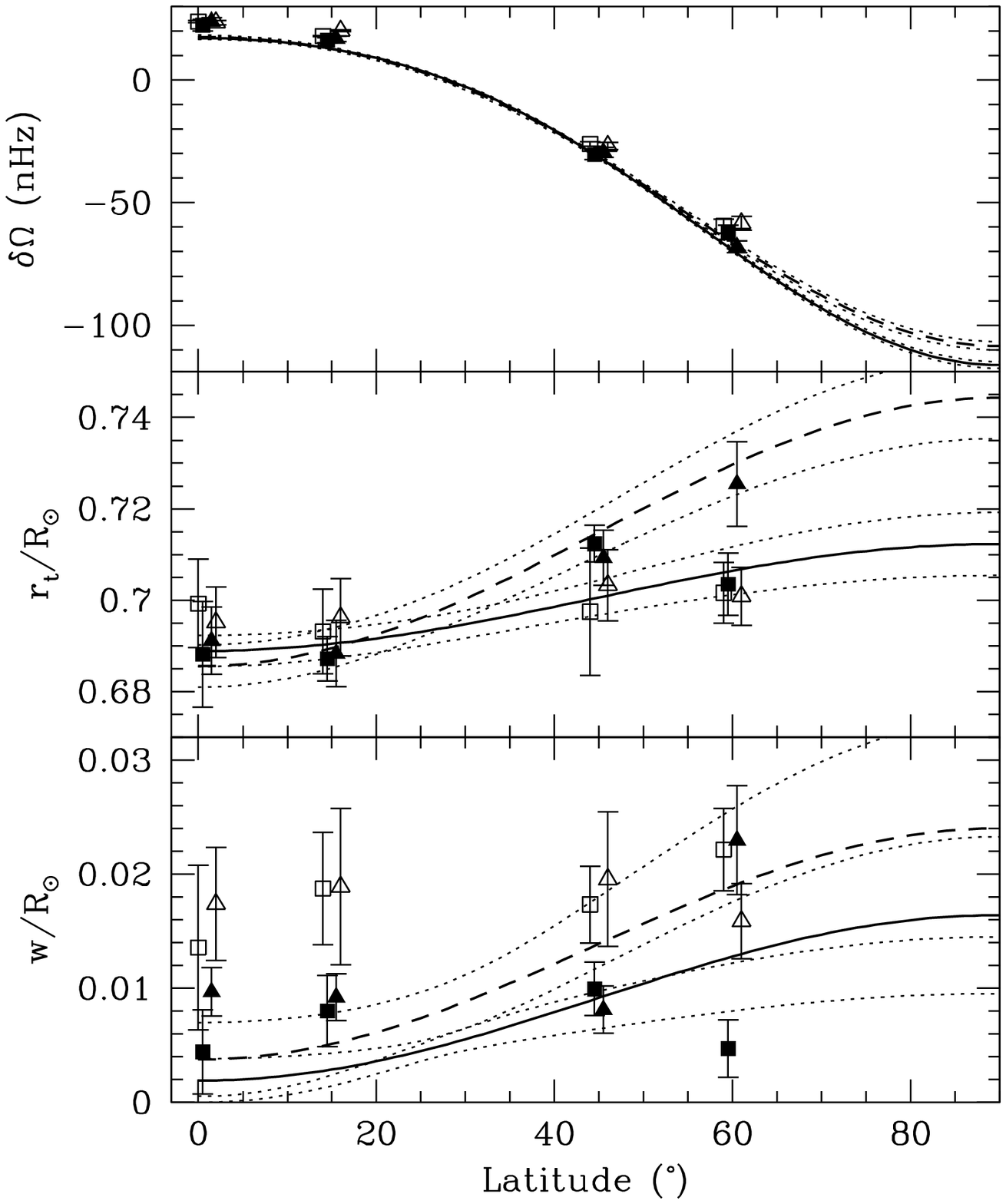}}
\caption{{\bf Figure 9.} Temporal average of tachocline properties
obtained from both GONG and MDI data. In each panel the continuous
and dashed lines show the results obtained using the 2d annealing
technique for GONG and MDI data respectively. The dotted lines
show the error estimates on these. The squares and triangles,
respectively, show the result obtained using GONG and MDI data.
The filled symbols represent results obtained using 1d annealing
technique, while the open symbols show results from the calibration
method. These symbols are shifted by $0.5^\circ$ or $1^\circ$ about the
true latitude for clarity.
}
\endfigure

\subsection{Latitudinal variation of the Tachocline}
In addition to  temporal variations, we can also study
latitudinal variations in the properties of the  tachocline.
For this
purpose we take the average of the tachocline properties
obtained from all the data sets.
Averaging over the 15 sets of non-overlapping GONG results obtained
 using the 2d annealing technique
we get the following results:
$$\eqalign{
r_t&=[(0.6936\pm0.0020)+(0.0047\pm0.0010)P_3(\theta)]R_\odot,\cr
w&=[(0.0048\pm0.0009)+(0.0029\pm0.0017)P_3(\theta)]R_\odot,\cr
\delta\Omega&=-(1.27\pm0.47)-(21.86\pm0.21)P_3(\theta)\cr
&\qquad-(3.44\pm0.11)P_5(\theta)\; {\rm nHz}.\cr}
\eqno\eqname\tachgong
$$
Similarly, the average over the 18 sets of MDI  2d annealing
results is:
$$\eqalign{
r_t&=[(0.6973\pm0.0028)+(0.0118\pm0.0013)P_3(\theta)]R_\odot,\cr
w&=[(0.0078\pm0.0023)+(0.0040\pm0.0023)P_3(\theta)]R_\odot,\cr
\delta\Omega&=-(1.17\pm0.68)-(21.47\pm0.28)P_3(\theta)\cr
&\qquad -(2.68\pm0.14)P_5(\theta)\; {\rm nHz}.\cr}
\eqno\eqname\tachgong
$$
These results are similar to those found by Antia et al.~(1998) using
a different data set from GONG.
It is clear that both sets of data show significant variation in
the position of the tachocline with latitude,
although the extent of variation
is quite different -- MDI results showing much more variation than
GONG results. The latitudinal variation in thickness is not
significant in either set.
There is clearly some systematic difference between the MDI and
GONG results, particularly, in the latitudinally varying component of
$r_t$. The origin of this difference is not clear. Other fitted
parameters are in reasonable agreement between the two sets of results.

\begintable{1}
\caption{\bf Table 1. \rm Mean properties of tachocline at different
latitudes}
{\tablet{8.5 true cm}{\hfil#\hfil&&\hfil$#$\cr
\tabmidrule
Latitude&\delta\Omega\hfil&r_t\hfil&w\hfil\cr
($^\circ$)&{\rm (nHz)}\hfil&(R_\odot)\hfil&(R_\odot)\hfil\cr
\tabmidrule
\phantom{0}0&21.18\pm0.27&0.6893\pm0.0023&0.0061\pm0.0012\cr
15&17.48\pm0.17&0.6899\pm0.0021&0.0061\pm0.0010\cr
45&-30.11\pm0.39&0.7077\pm0.0021&0.0108\pm0.0012\cr
60&-67.27\pm0.58&0.7093\pm0.0027&0.0135\pm0.0015\cr
\tabmidrule}}
\endtable

Figure 9 compares the time-averaged results of the tachocline
properties obtained using different techniques and data sets 
as a function of latitude.
It is clear that all results show that the position of the tachocline
moves outwards with latitude. In fact, this trend can be seen in
the individual results shown in Fig.~3 also.
The latitudinal variation in width is not clear, but the results
tend to suggest that the thickness increases with latitude.
Though it is possible that this increase may be due to increase in
errors with latitude.
The jump in rotation rate across the tachocline, $\delta\Omega$,
of course, shows a pronounced latitudinal variation, which is well
known.
In order to get better idea of latitudinal variation
we take weighted average over all six
measurements at each latitude and the results are shown in Table~1.
Thus the increase
by $0.020R_\odot$ in $r_t$ with latitude is significantly larger
than the error estimates, while the increase in half-width by
$0.006R_\odot$ is only marginally significant.
The shift in tachocline position with latitude is comparable to that
found by Charbonneau et al.~(1999). Thus our results also support
the conclusion that the tachocline is prolate.

\subsection{Latitudinal variation in the depth of the convection zone
and extent of overshoot}

\begintable{2}
\caption{\bf Table 2. \rm Mean position of base of the convection zone
and extent of overshoot at different latitudes}
{\tablet{8.5 true cm}{\hfil#\hfil&&\hfil$#$\cr
\tabmidrule
Latitude&r_d\hfil&A\hfil&\tau\hfil\cr
($^\circ$)&(R_\odot)\hfil&(\mu {\rm Hz})\hfil&(\rm s)\hfil\cr
\tabmidrule
Mean&0.71336\pm0.00002&0.716\pm0.010&2304\pm3\phantom{0}\cr
\noalign{\medskip}
\phantom{0}0&0.71333\pm0.00010&0.889\pm0.095&2330\pm24\cr
15&0.71338\pm0.00008&0.758\pm0.050&2314\pm12\cr
30&0.71334\pm0.00008&0.905\pm0.055&2273\pm14\cr
45&0.71324\pm0.00008&0.761\pm0.064&2329\pm15\cr
60&0.71359\pm0.00013&0.602\pm0.165&2310\pm97\cr
75&0.71312\pm0.00080&&\cr
\tabmidrule}}
\endtable

Having found the tachocline to be prolate, it is of interest to
check whether the base of the convection zone has the same shape.
The relative position of the base of the convection zone with respect
to the tachocline plays a crucial role in theoretical models to
explain the formation of the tachocline.
To determine the depth of the convection zone at different latitudes we
construct frequency differences between a solar model and the Sun
using Eq.~\delnu\ and apply the same technique as that used in
Section 4.1 to determine the position of the base of the convection
zone.  Once again we  time-average the results to 
determine the latitudinal variation more precisely than what we would be
able to with just one set of data.
The results at a few different latitudes are shown in Table~2.
The first row in this table is the result for the spherically
symmetric case.
These results do not show any clear indication of latitudinal variation
in the depth of the convection zone and we can put an upper
limit of about $0.0005R_\odot$ on possible variations. Nevertheless, the
results appear to indicate that there is a marginally significant
increase in $r_d$ around a latitude of $60^\circ$. Interestingly,
Antia et al.~(2000a),  in their inversion results for the  aspherical
component of the solar sound speed, find an excess over the spherically symmetric
value of the  sound speed in the convection zone  around this latitude.
It is possible that this feature extends
up to the base of the convection zone and gives rise to this small
excess at $60^\circ$ in $r_d$. More data are needed to confirm
the departure from sphericity of the base of the convection
zone. In any case, this difference is two orders of magnitude less
than the variation claimed by Gough \& Kosovichev (1995). We can
certainly rule out a variation of $0.02R_\odot$, which is what is seen 
for the tachocline position. Thus it is clear that while the tachocline has
a prolate shape, the base of the convection zone is more or less
spherical.

We also attempt to determine the latitudinal dependence of the extent of
overshoot below the convection zone, by studying the amplitude of
the oscillatory component in the fourth differences of the
frequencies with appropriate combinations of even splitting coefficients
added. In this case, since the errors in individual data sets are
too large to isolate the oscillatory component in frequencies, we
take average frequencies over all available data sets in GONG or MDI
for each latitude and then study the oscillatory component in these.
The results are shown in Table~2. Even after averaging over all
sets, the errors are still substantial and it is not possible to
get any meaningful results at high latitudes. That is why we haven't
included the results at $75^\circ$ latitude.
There is no clear latitudinal dependence
in either the amplitude, $A$ of the oscillatory signal or its
frequency, $\tau$. The amplitude is consistent with that of a
model without any overshoot. The amplitude is generally larger than what is
seen for the mean spherically symmetric data, but the increase may
not be significant as the amplitude tends to be overestimated when
the data have substantial error (Basu 1997).
Thus we do not find any evidence for latitudinal
variation in the extent of overshoot below the solar convection zone.

\section{Discussion}
Using data from GONG and MDI covering the period 1995--2000 we have
tried to determine temporal variations in structure and dynamics near
the base of the solar convection zone. This region is widely believed
to be the seat of solar dynamo.
We fail to find any significant signal of time variation, solar-cycle
related or otherwise, in either the structure, or the  dynamics of 
the base of the solar convection zone. The depth of the convection zone 
does not show
any significant change with time, nor does the extent of overshoot.
Any systematic variation in the convection-zone depth  should be 
small, with the estimated change being $28\pm25$ km.

We do not find any convincing evidence of change in the dynamics near
the solar convection zone base either.
We have used three different forward modelling techniques to study
the characteristics of the solar tachocline.
Neither the position, nor the  thickness
of the tachocline show any temporal variation, while the
result for the 
jump in the rotation rate across the tachocline depends slightly 
on the technique used in studying the tachocline.
The results obtained by the calibration technique show a mild increase in 
the jump 
with time at low latitudes, the increase being  comparable to the estimated
error in the results. The annealing
techniques, on the other hand,  do not show any systematic increase. 
The 2d annealing
results show some oscillatory behaviour. There is a weak periodic
signal in both MDI and GONG data for $\delta\Omega_1$, the
spherically symmetric component of
the jump in rotation rate across the tachocline and $\Omega_c$, the
spherically symmetric component of rotation rate in the radiative interior.
The Fourier transforms of $\delta\Omega_1$ or $\Omega_c$ obtained using only
the splitting coefficient $a_1$
show peaks around a frequency of 0.8--0.9 yr$^{-1}$
which are at $1\sigma$ level. The frequency resolution of the
transforms is
0.25 yr$^{-1}$, thus  these peaks are close to a period of 1 year, which
may be expected from the orbital motion of the Earth.
The significance of these peaks is even more unclear since  the 2d annealing fits
show  an anti-correlation between the parameters $\delta\Omega_1$
and $\Omega_c$. This is most probably caused by  
the difficulty in distinguishing between these parameters in $a_1$
for the model of tachocline used to fit the observed data.

We do not find any convincing evidence of 1.3 year periodicity
found by Howe et al.~(2000). 
A closer look at their figure shows that
the periodic behaviour is seen clearly in only the GONG results
and is not clear in the results obtained using MDI data. 
As such the
significance of this periodicity is questionable. 
We do not find any
significant periodicity even in the GONG data. We find a
marginally significant periodicity in the spherically symmetric
component of rotation rate below the tachocline. This is quite
different from strong latitudinal variation claimed by Howe et
al.~(2000). 
The spherically symmetric component of rotation rate
is determined using  the splitting coefficient $a_1$, which has to be
corrected for the effects of 
the orbital motion of Earth. The correction from synodic to sidereal
rotation rate depends on the Earth's orbital motion, and there may be
some error in applying this correction.  Thus one may expect a periodicity 
of 1 year in the observed values of $a_1$. The limited frequency 
resolution of current data sets may cause the peak in the Fourier spectrum
to be shifted to slightly  different values of the period.
Were this to be true,
 one would expect this effect to give oscillations in $\Omega$
which are independent of depth. However, we find a depth
dependence in the oscillations, and moreover, two independent data sets appear
to agree with each other, suggesting a solar origin of the signal. 
It is quite possible that the periodic signal arising from
orbital effects depends on the degree, $\ell$,  in which case 
a depth-dependence is expected. If this signal indeed represents real
oscillations on the Sun, then it is not clear if it is related to
solar activity or solar dynamo. Origin of such a periodicity will
be difficult to understand theoretically. A periodicity
with period around 1 year is also seen in the frequencies of
f-modes (Antia et al.~2000, in preparation) which are confined
to layers just below the solar surface.

The GONG and MDI  data can also be used to study latitudinal variations 
in the properties of the 
tachocline. By time-averaging the results we
find that the tachocline is prolate, with the difference between the
tachocline position at $0^\circ$ and that at $60^\circ$ latitude being about
$(0.020\pm0.003)R_\odot$. This is
in agreement with results obtained by (Charbonneau et al.~1999).
There is also some increase in thickness of the tachocline with
latitude, by about $(0.006\pm0.002)R_\odot$.  This increase is
less significant, though still at the $3\sigma$ level.
We do not find any clear evidence for latitudinal variations in the depth of
the convection zone or the extent of overshoot below the convection
zone. However, there is a small increase in $r_d$, the position of
the base of the convection zone around the latitude of $60^\circ$
by $(0.0002\pm0.0002)R_\odot$ as compared to the mean value.
The significance of this departure is not clear, but it may be noted that
Antia et al.~(2000a) found a significant excess in sound speed around
this latitude, which is concentrated around a radial distance of
$0.92R_\odot$. It is quite possible that this has some effect near
the base of the convection zone, causing a small difference in $r_d$.
Combining the results on latitudinal variation of the tachocline and
the depth of the convection zone, we can conclude that around
the solar equator, the  bulk of the
tachocline is below the base of the convection zone, but at higher
latitude a substantial part of the tachocline  moves into the convection zone.

The latitudinal dependence of the  position of the tachocline will have an important
bearing on the theoretical models of the tachocline. Many models
 are based on the assumption that the tachocline is located
in the radiative region. Canuto~(1998), however,  suggested that buoyancy and
vorticity inside the convection zone can account for the small thickness
of the tachocline. This model may still be admissible since, at least at
high latitudes, about half of the tachocline is inside the convection
zone. The thickness of $0.05R_\odot$ for the tachocline as calculated by Canuto,
is much larger than what we find, but there is considerable
ambiguity in the definition of thickness, since it depends on the actual
model of the rotation rate within the tachocline. Furthermore,
proper calculations using
Canuto's model may even give a different value. So this discrepancy
in thickness may not rule out the model.
Gough \& McIntyre~(1998) have proposed that a magnetic field
may provide the required horizontal transport to keep the thickness low.
However, their estimate of the magnetic field depends on ninth power of the
thickness. Considering the ambiguity in the definition of the thickness,
it is difficult to draw any conclusions
about the validity of this model, but if we use the equatorial
value of the thickness as estimated by us, the required magnetic
field turns out to be of the order of 200 G.
The physical origin
of the prolateness of the tachocline is even more difficult to
explain, particularly, since the convection zone itself does not appear
to be prolate. A difference between the shape of the tachocline and 
that of the convection
zone base is rather difficult to understand theoretically.
It is quite likely that magnetic field, which is estimated
to be $10^5$ G (D'Silva \& Choudhuri 1993)
in these layers, plays an important role in determining
the shape and thickness of the tachocline as well as its relative
position with respect to the convection zone base.

\section*{Acknowledgments}

This work utilises data obtained by the Global Oscillation
Network Group (GONG) project, managed by the National Solar Observatory,
which is
operated by AURA, Inc. under a cooperative agreement with the National
Science Foundation.
The data were acquired by instruments operated by the Big
Bear Solar Observatory, High Altitude Observatory,
Learmonth Solar Observatory, Udaipur Solar Observatory,
Instituto de Astrofisico de Canarias, and Cerro Tololo
Interamerican Observatory.
This work also utilizes data from the Solar Oscillations
Investigation / Michelson Doppler Imager (SOI/MDI) on the Solar
and Heliospheric Observatory (SOHO).  SOHO is a project of
international cooperation between ESA and NASA.

\bye